\documentclass[journal]{IEEEtran}

\newcommand{\subparagraph}{} %
\usepackage{gensymb}

\usepackage[utf8]{inputenc}
\usepackage{titlesec}
\usepackage{xcolor}
\newsavebox{\imgbox}
\usepackage[bottom]{footmisc}
\usepackage[pdftex]{graphicx}
\usepackage[scaled]{helvet}
\usepackage{cite}
\usepackage{epstopdf}
\usepackage{amssymb}
\usepackage{bigints}
\usepackage{xfrac}
\usepackage{amsmath}
\usepackage{graphicx}
\usepackage{multirow}
\usepackage[none]{hyphenat}
\usepackage{float}
\usepackage[caption=false]{subfig} 
\usepackage{dblfloatfix}

\usepackage{t1enc}
\usepackage{times}

\pdfoutput = 1

\begin{document}



\title{Adaptive Distributed Transceiver Synchronization Over a 90 Meter Microwave Wireless Link}

\author{Serge R. Mghabghab,~\IEEEmembership{Student Member,~IEEE}, Anton Schlegel,~\IEEEmembership{Student Member,~IEEE},\\and Jeffrey A. Nanzer,~\IEEEmembership{Senior Member,~IEEE}
	
	\thanks{Manuscript received 2020.}
	\thanks{
		This work was supported in part by the Office of Naval Research under Grant N00014-17-1-2886 and in part by the Defense Advanced Research Projects Agency under Grant N66001-17-1-4045. The views, opinions, and/or findings contained in this article are those of the author and should not be interpreted as representing the official views or policies, either expressed or implied, of the Defense Advanced Research Projects Agency or the Department of Defense.
		\textit{(Corresponding author: Jeffrey A. Nanzer.)}
	}
	\thanks{The authors are with the Department of Electrical and Computer Engineering, Michigan State University, East Lansing, MI 48824 USA (email: \{mghabgha, schleg19, nanzer\}@msu.edu).}
}

\markboth{IEEE}%
{Shell \MakeLowercase{\textit{et al.}}: Bare Demo of IEEEtran.cls for Journals}

\maketitle



\begin{abstract}
We present an adaptive approach for synchronizing both the phase and frequency of radio-frequency transceivers over long-range wireless links to support distributed antenna array applications. To enable distributed beamforming between separate wireless nodes, the oscillators in the transceivers must operate at the same frequency, and their phases must be appropriately aligned to support phase-coherent beamsteering. Based on a spectrally-sparse waveform, a self-mixing circuit, and an adaptive control loop, we present a system capable of synchronizing the RF oscillators in separate transceivers over distances of nearly 100 m. The approach is based on a spectrally-sparse waveform for joint inter-node ranging and frequency transfer. A frequency reference is modulated onto one signal of a two-tone waveform transmitted by the primary node which is demodulated and used to lock the oscillator of the secondary node. The secondary node retransmits the two-tone signal which the primary node uses for a high-accuracy range measurement. From this range, the phase of the two transceivers can be aligned to support beamforming. We furthermore implemented an adaptive phase control approach to support high-accuracy phase coordination in changing environmental conditions. We demonstrate continuous high accuracy links over a 90 m distance in an outdoor environment for durations up to seven days, demonstrating sufficient phase coordination in changing weather conditions to support distributed beamforming at frequencies up to 3 GHz.


\end{abstract}



\IEEEpeerreviewmaketitle

\section{Introduction}


\IEEEPARstart{I}{mproving the performance} of wireless remote sensing and communications entails enhancing the size, weight, cost, power consumption, and quality of these systems. 
In platform-centric systems there is, however, a trade-off between the possible system enhancement and the cost and size of the system. These limitations can be lifted by focusing on distributed wireless subsystems in place of single-platform systems, which can achieve high performance while mitigating the use of expensive bulky systems. 
Among the approaches to implementing multi-platform electromagnetic coordination, the most relevant for distributed wireless systems are Multiple Input Multiple Output (MIMO) approach \cite{1353475, 8359454, 8660656}
or distributed beamforming \cite{7880558, 4202181, 9173801}. 
Coherent distributed arrays are a specific category of distributed wireless subsystems that can achieve distributed beamforming by coordinating the relative phases of the nodes in the array. These arrays have a wide set of applications including wireless sensing, communications, and distributed microwave imaging, and can be implemented on small radios, satellites, airplanes, ground vehicles, unmanned aerial vehicles, and other platforms. However, the benefits of distributed beamforming come at a cost, which is the sophisticated coordination requirements. To ensure coherent operation among all distributed nodes, all elements need to be frequency locked to ensure that all oscillators are operating at the same frequency \cite{6949699, 9028079, 7218555, SergeIMSBeamform}; time aligned to ensure proper alignment of information at destination \cite{5893884, 7421334, 8678474, 6624252}; and phase aligned in order to have coherent summation of the signals at the destination \cite{5670903, 6488994, 7460546, 4542555, 7795367}.

Distributed phased array architectures can be divided into two main groups: closed-loop and open-loop. In closed-loop architectures, feedback signals from the targeted location are used in the inter-node synchronization process \cite{5670903, 6488994, 7460546, 4542555, 7795367}. This approach is easier to implement in comparison to open-loop, however it is applicable only to a limited set of applications, since it is not possible to use this set of architectures whenever the targeted location cannot or does not cooperate with the transmitters. Thus, sensing applications and communications link initialization are generally not supported. On the other hand, open-loop architectures support a wide application space since the node synchronization is performed by relying only on inter-node synchronization signals, and can beamform without any feedback from the destination. Open-loop architectures are not constrained to any space and can be used in a flexible manner as long as proper synchronization withing the array is achievable. Among the synchronization signals, frequency locking and phase alignment have the most stringent requirements \cite{nanzer2017open}, and are the focus of this work.
In other works, wireless frequency synchronization has been achieved in various ways, including using coupled-oscillators \cite{ponton2017stability}, optically-locked voltage controlled oscillators\cite{yang2014picosecond}, and trading of data packets \cite{9028079, morelli2007synchronization, 8742232}, among others. These techniques, however, are either not feasible for long inter-node separations, have discretization errors which contribute to significant phase errors, allow small frequency drifts between the frequency update intervals, or produce phase shifts which are not possible to track in a dynamic array setting. 

In this work we present a framework for adaptive, long-range phase synchronization of distributed transceivers for coherent distributed beamforming applications. We build on our prior work on wireless frequency locking using a unique physical-layer approach based on an adjunct self-mixing circuit receiving spectrally sparse waveforms to align the frequency of the local oscillator of the transceiver to that of a distant transceiver \cite{8889331, SergeTMTTBeamform}. We combine this with a high-accuracy inter-node ranging technique that similarly uses spectrally-sparse waveforms to adjust the oscillator phases to enable phase-coherent beamforming. We demonstrate distributed phase synchronization between two microwave transceivers separated by 90 m operating in an outdoor environment over a period of days. A cooperative ranging approach was performed to boost the received signal-to-noise ratio (SNR) of the ranging signals. Furthermore, since various weather and interference conditions, as well as antenna vibrations, can affect the SNR in an unpredictable fashion, adaptive ranging was implemented using a proportional-integral (PI) controller to maintain a desired ranging accuracy. 
Our experiments demonstrate the ability of the adaptive, long-range phase synchronization approach to support distributed beamforming at frequencies up to 3 GHz in the presence of changing environmental conditions.

\section{Phase Synchronization in\\Distributed Wireless Systems}

\begin{figure}[t!]
	\centering
	\includegraphics[width=\linewidth]{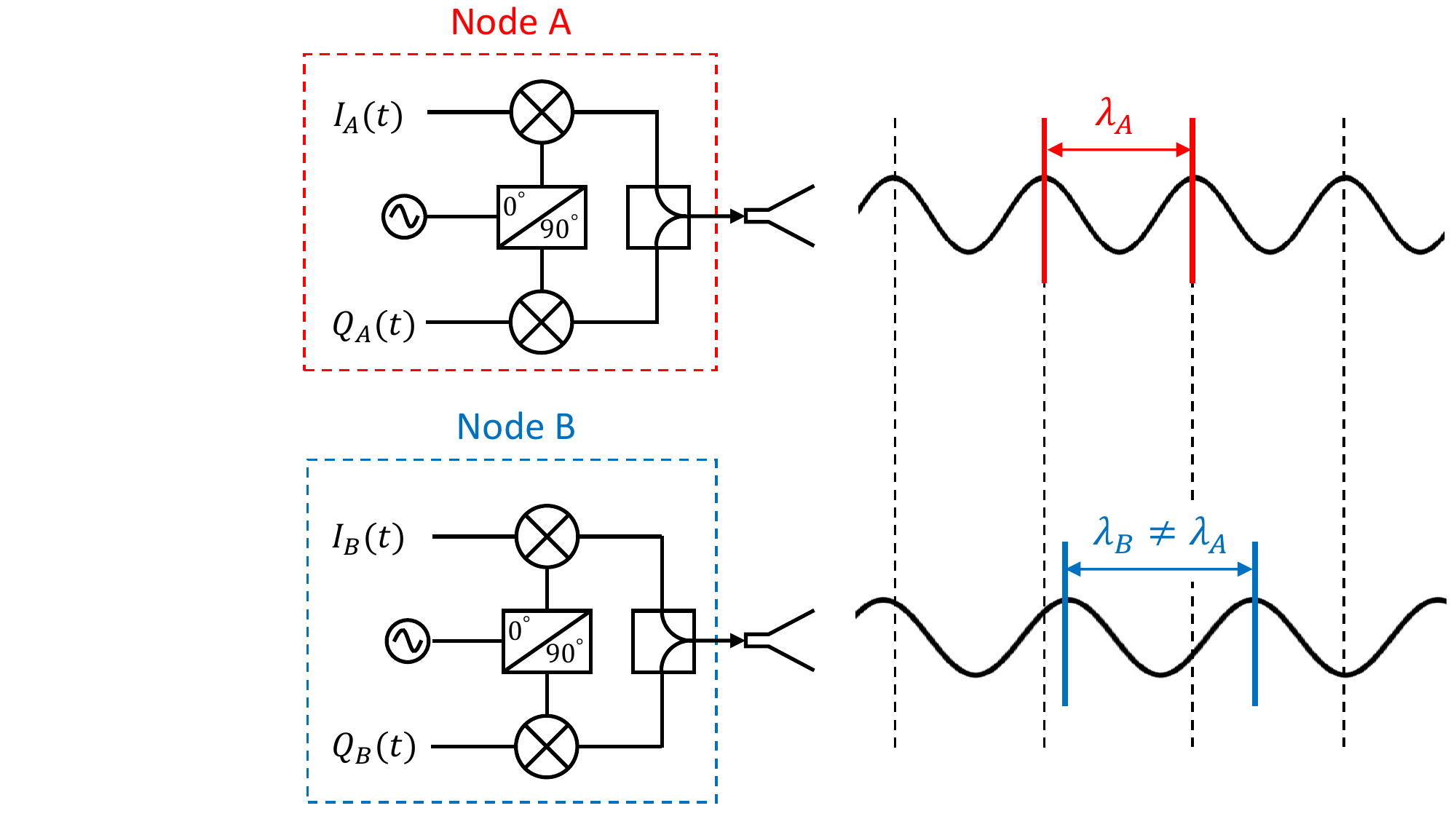}
	
	(a)
	
	\includegraphics[width=\linewidth]{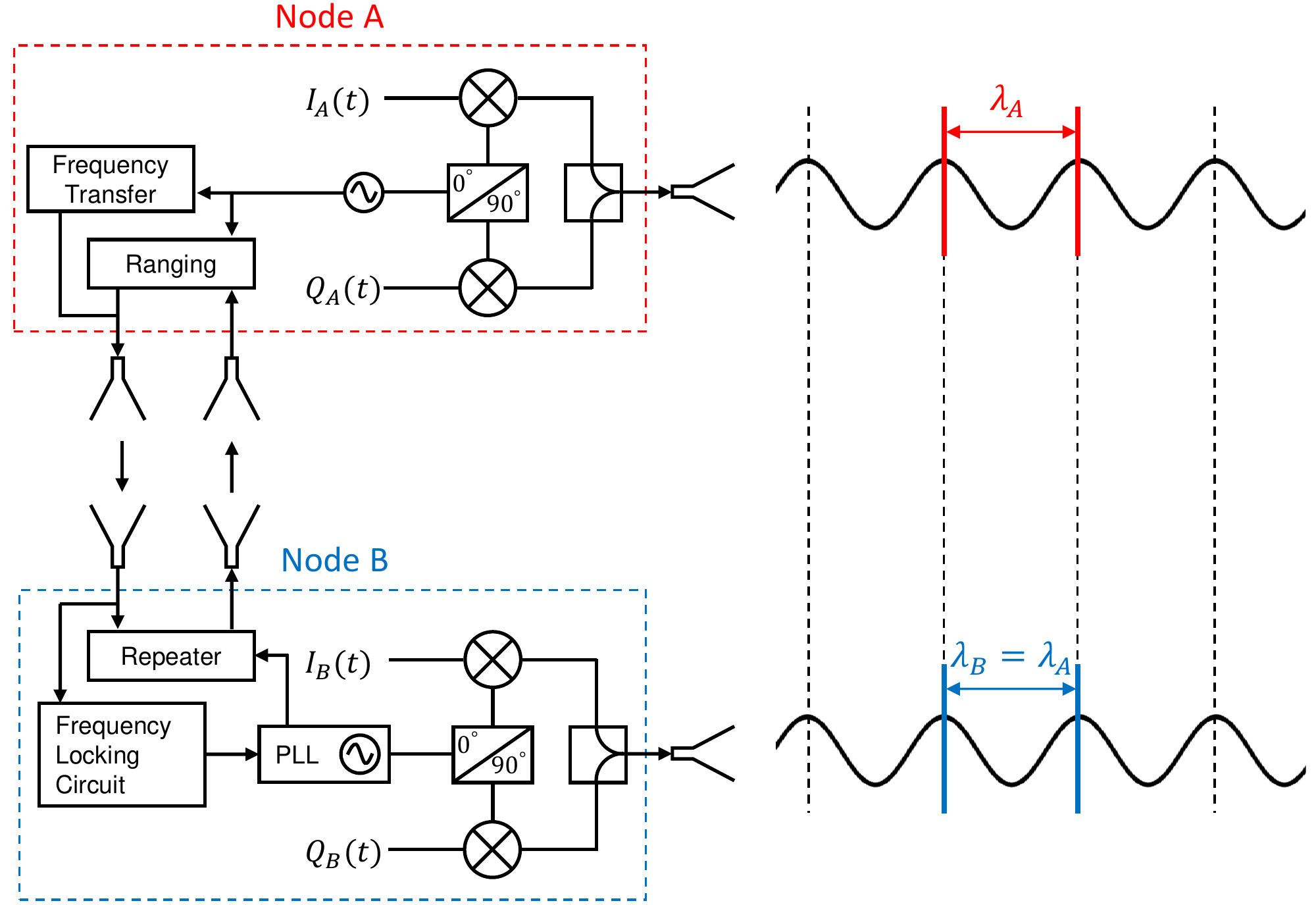}
	
	(b)
	
	\caption[Optional caption]{(a) Without synchronization of their oscillator frequency and phases, intrinsic phase noise and frequency drift result in separate transmitters emitting signals at slightly different frequencies and different phases, yielding low coherent gain at the destination. (b) By wirelessly synchronizing the frequencies and phases of the oscillators, distributed transmission can be accomplished at the same frequency with appropriate phase alignment to support coherent beamforming. In this work we implement phase and frequency alignment through adaptive high-accuracy ranging and wireless frequency transfer.}
	\label{fig:OV-1}
\end{figure}

The principal objective of distributed phase synchronization is to ensure that separate wireless transceivers transmit signals at the same frequency, and in a phase-aligned state such that a coherent phase front is maintained in the desired beamforming direction. Fig. \ref{fig:OV-1}(a) shows an example of two transceivers, each consisting of baseband I/Q data, a local oscillator (LO), a $90\degree$ hybrid providing the in-phase and quadrature oscillator signals, and upconverters. In the case where the local oscillators on each node are operated independently, the frequencies will naturally drift relative to one another, even if the oscillators are designed at the same nominal frequency. Furthermore, the relative phase relationship between the oscillators may be dynamic. The result is the transmission of two signals that will rotate through in-phase and out-of-phase relationships, resulting in low coherent gain. 

Our architecture corrects for the frequency and phase differences using the topology shown in Fig. \ref{fig:OV-1}(b). Node A transmits a signal that is modulated by the frequency reference derived from it LO to node B. This signal is received by node B, which enables two functions. The first is to retransmit the receive signal back to node A, whereby node A estimates the range between the two to implement phase-coherent beamsteering. Second, node B demodulates the frequency references from node A and inputs it to a phase-locked loop (PLL) which locks the LO of node B to that of node A. With this approach, the frequencies of the two transceivers will be locked, and their relative phases can be adjusted in response to any changes in distance between them to maintain a coherent phase front. 
The errors for the relative inter-node localization need to be in the order of small fractions of the transmitted signal wavelength in order to ensure coherent operation. Furthermore, the ability to estimate the range between the two nodes is dependent on the received SNR, thus any environmental changes which may degrade the SNR will impact the range estimate, leading to greater errors and degraded beamforming performance. In the following sections, we describe in detail the inter-node ranging approach and the wireless frequency transfer technique, as well as an adaptive approach to maintaining high-accuracy range measurements in changing conditions. Following this, we demonstrate the adaptive ranging implementation in an outdoor experiment operating over 90 m and continuously over time frames of days.

\begin{figure}[t!]
	\centering
	\includegraphics[width=\linewidth]{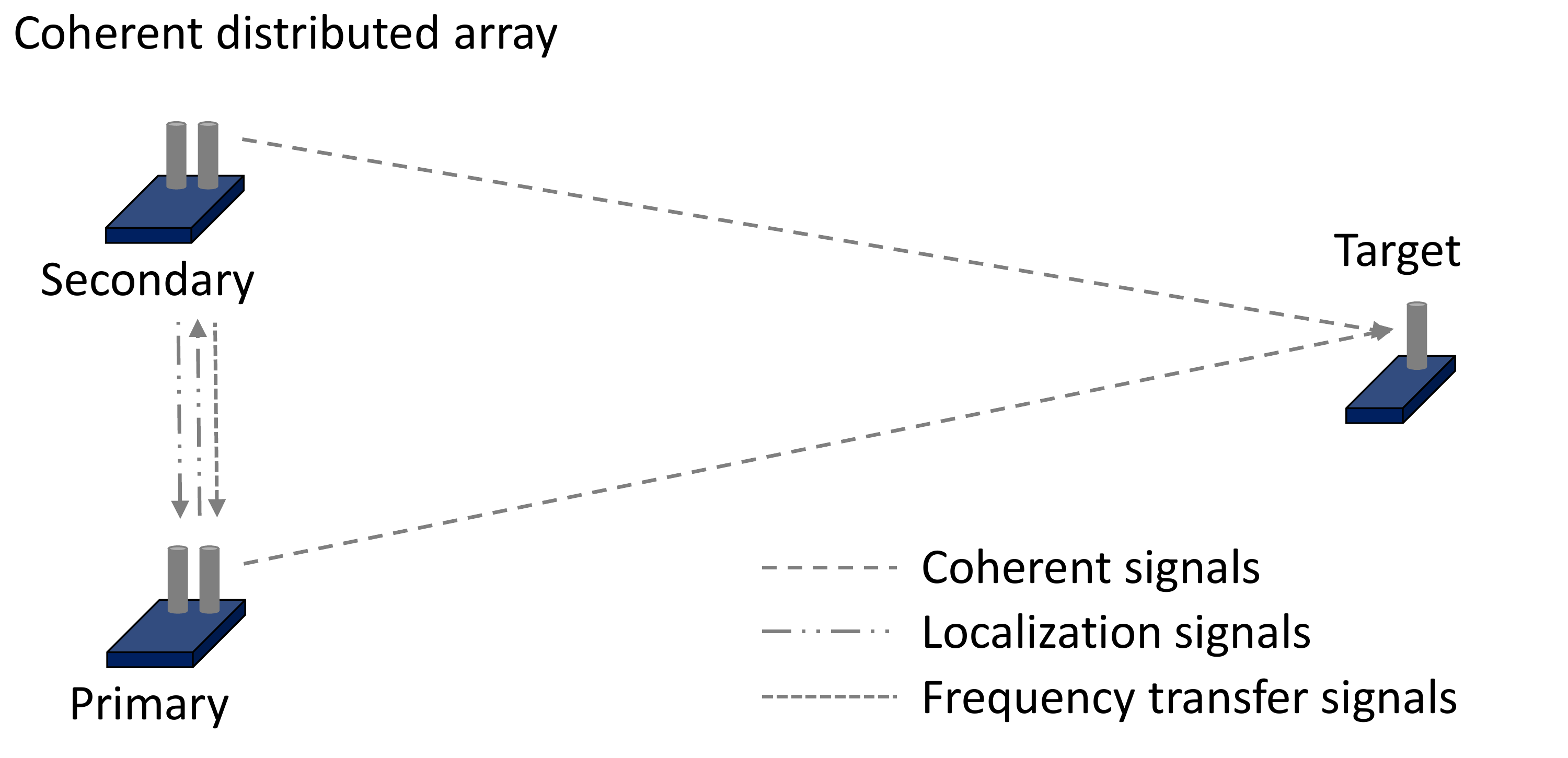}
	\caption[Optional caption]{A two-node open-loop coherent distributed antenna array. For coherent operation, it is necessary to ensure proper localization and frequency synchronization.}
	\label{fig:CDA}
\end{figure}


Any realistic system will impart errors, and the errors in phase synchronization stemming from the inter-node range estimates will impact the beamforming ability. To determine what error can be tolerated, we evaluate the coherent gain $G_c$, which is calculated using
\begin{equation}\label{eq:Gc}
G_c = \frac{\left|s_r s_r^*\right|}{\left|s_i s_i^*\right|},
\end{equation}
where $s_r$ represents the summation of the beamformed signals in the presence of errors, and $s_i$ represents the ideal, error-free summation of transmitted signals at the targeted location. The metric $G_c$ is equal to 1 when perfect phase correction is implemented.

The ideal summation of the received signals from $N$ tansmitting nodes can be expressed as
\begin{equation}\label{eq:si}
s_i(t) = \sum_{n=1}^{N} h_n A_n(t) e^{j2\pi f_c t},
\end{equation}
where $h_n$ is the complex valued coefficient for the $n^{\mathrm{th}}$ propagation channel, $A_n$ represents the amplitude of the $n^{\mathrm{th}}$ signal, and $f_c$ is the center frequency of the signal.
The beamformed signals with errors $s_r$ can be represented as
\begin{equation}\label{eq:sr}
s_r(t) = \sum_{n=1}^{N} h_n A_n(t) e^{j\left[ 2\pi f_c t + \Delta \phi (n) - \Delta \phi_{e} (n)  +\delta\phi(n) +\phi_0(n) \right]},
\end{equation}
{where $\Delta \phi (n) = \tfrac{d(n)}{c}\sin\theta$ represents the phase shift necessary to steer a beam towards $\theta$, $\Delta \phi_{e}(n)~=~\tfrac{d(n) + \delta_d(n)}{c}\sin\theta$ represents the estimated phase correction in a dynamic system with $d(n) + \delta_d(n)$ the estimated inter-node distance with error $\delta_d(n)$. The term $\delta\phi(n)$ represents the phase error due to clock misalignment and frequency drift, and $\phi_0(n)$ is the residual error from the calibration of the initial phase shifts; this calibration is performed upon initialization.}

\begin{figure}[t!]
	\centering
	\includegraphics[width=\linewidth]{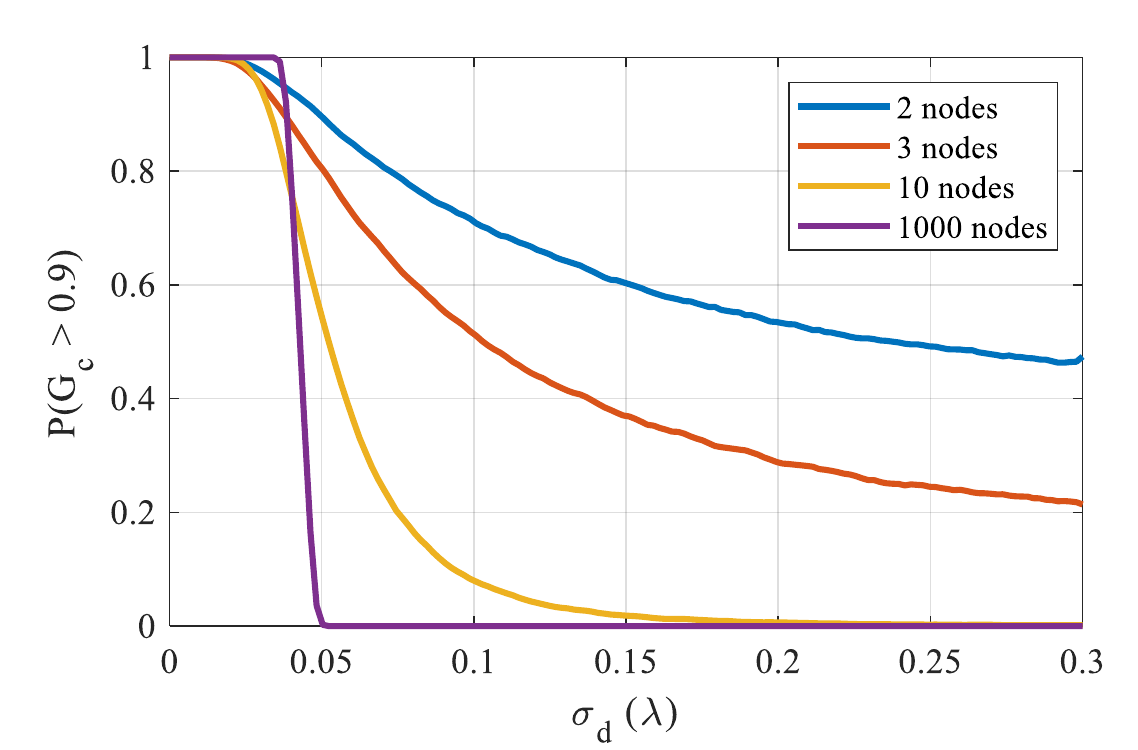}
	\caption[Optional caption]{Probability of the coherent gain exceeding 0.9 versus the ranging standard deviation in the case where wireless frequency synchronization was achieved using the adjunct self-mixing circuit \cite{SergeAPS2020RangingRequirements}.}
	\label{fig:rangingReq}
\end{figure}

The coherent gain is affected by multiple factors such as phase noise, instantaneous phase and frequency errors, time alignment, angle estimation, and range estimation. In this work we study the effect of range accuracy on the coherent gain. Ranging is used to align the phases of all the transmitted signals, making it an important and challenging task, since any small error in the range estimates can lead to reduced coherence. 
We evaluate the probability to achieve coherent gain above a certain threshold n order to determine the desired ranging accuracy or standard deviation for our system; this probability is expressed as
\begin{equation}\label{eq:PGc}
Y = P\left(G_c \geq X\right),
\end{equation}
where $0 \leq X \leq 1$ is the threshold of interest.
10,000 Monte Carlo Simulations were performed in \cite{SergeAPS2020RangingRequirements} to evaluate the probability to achieve at least 90\% coherent gain ($P\left(G_c \geq 0.9\right)$). In these simulations, the transmission angle of the coherent signal, the distance separating the frequency synchronization antennas, and the distance separating the nodes transmitters were randomized. In this study, no initial calibration errors were considered; $\phi_0(n) = 0$. Fig. \ref{fig:rangingReq} shows the smoothed results for 2, 3, 10, and 1000 nodes. For a two-node system, like that used in this paper, to have 90\%, 80\%, and 70\% probability for achieving at least 90\% coherent gain, the ranging standard deviation $\sigma _d$ needs to be at most $0.0495\lambda$, $0.0725\lambda$, and $0.1040\lambda$ respectively, where $\lambda$ is the wavelength of the coherent signal. These metrics will be used later in the paper to analyze the achieved ranging standard deviation.

\section{High-Accuracy Ranging for Phase Alignment}

To appropriately phase-align the transmitted signals to the desired direction for beamforming, the range between the transmitting node and a reference point (e.g. the primary node) must be accurately estimated.
In \cite{weiss1983fundamental, weinstein1984fundamental, 7801084}, it was shown that two-tone signals achieve  optimal ranging accuracy, albeit at the expense of increased ambiguities. However, for cooperative systems such as between nodes in a distributed array, the ambiguities are mitigated since the cooperating node retransmits the incident ranging signal, thereby providing a point-like impulse response. For multiple cooperative nodes with active repeaters, two-tone stepped frequency waveforms were developed in \cite{9057428} to allow high accuracy ranging in a multi-node system. Selection of the appropriate ambiguity lobe for unambiguous ranging has been addressed in \cite{hodkin2015microwave, 7801084}, where short pulses were transmitted between the ranging signals; these pulses helped locating the main peaks at the output of the matched filters.

For moderate to high SNR, the minimum achievable variance for the time delay estimates is governed by the Cramer Rao lower bound (CRLB)
\begin{equation}\label{eq:CRLB}
\sigma_t^2 = \frac{1}{\beta ^2 \frac{2E}{N_0}},
\end{equation}
where $\beta ^2$ is the mean-squared bandwidth of the waveform, $E$ is the signal energy, and $N_0$ is the noise power spectral density. The ratio $\frac{E}{N_0}$ is proportional to the post-processing SNR: $\frac{E}{N_0} = TB \cdot SNR$, where $T$ is the pulse duration, and $B$ is the bandwidth. The mean-squared bandwidth is given by  
\begin{equation}\label{eq:Beta2}
\beta ^2 = \frac{\int\limits_{- \infty}^\infty \left( 2 \pi f\right)^2 \left| S(f) \right|^2 df}{\int\limits_{- \infty}^\infty  \left| S(f) \right|^2 df} ,
\end{equation}
where $S(f)$ is the Fourier transform of the temporal signal.

A two-tone signal given by
\begin{equation}\label{eq:S(t)}
S(t) =  \left(e^{-j2 \pi t \delta f} + e^{j2 \pi t \delta f}\right) e^{j2 \pi t f_c},
\end{equation}
where $\delta f$ is one-half of the separation of the two tones, has a larger mean-squared bandwidth than other waveforms, yielding better accuracy. The mean-squared bandwidth for the two-tone waveform can be derived as 
\begin{equation}\label{eq:Beta2_2}
\beta ^2 = \left( 2 \pi \delta f \right) ^2 + \left( 2 \pi f_c \right) ^2.
\end{equation}
At baseband ($f_c = 0$), the CRLB for the ranging standard deviation is obtained from
\begin{equation}\label{eq:CRLB_r}
\sigma_{r} = \frac{c}{8\left( \pi \delta f \right)^2 \sqrt{ \frac{2E}{N_0}}},
\end{equation}
where $c$ is the speed of light. Note that $\sigma_{r}~=~c/2\sigma_{t}$, since the time delay is obtained from the two-way delay while the range estimates consider the one-way path factor.


The steps used in this experiment to perform accurate ranging measurements are described below:
\begin{enumerate}
	\item The frequencies of the two-tone ranging pulse along with a disambiguation pulse are designed to achieve the desired accuracy and then they are transmitted.
	\item Once the transmitted pulses are captured, amplified, and retransmitted by the targeted node, they are captured by the radar and matched filtered to improve their SNR.
	\item The main peak that belongs to the matched filtered output of the disambiguation pulse is located, afterwards the closest peak that belongs to the matched filtered output of the two-tone pulse is selected. The refined version of the selected peak is used to estimate the range of the target.
	\item The discretized lobe that contains the peak of interest is selected and interpolated using a spline interpolation to improve the estimation accuracy.
	\item The range of the targeted node is calculated from the time delay estimates. $N$ range estimates can be averaged to obtain a more accurate range estimate, producing a decrease in the standard deviation of the estimates by a factor of  $\sqrt{N}$.
\end{enumerate}

\begin{figure}[t!]
	\centering
	\centering%
	\subfloat[]{%
		\centering
		\includegraphics[width=\linewidth]{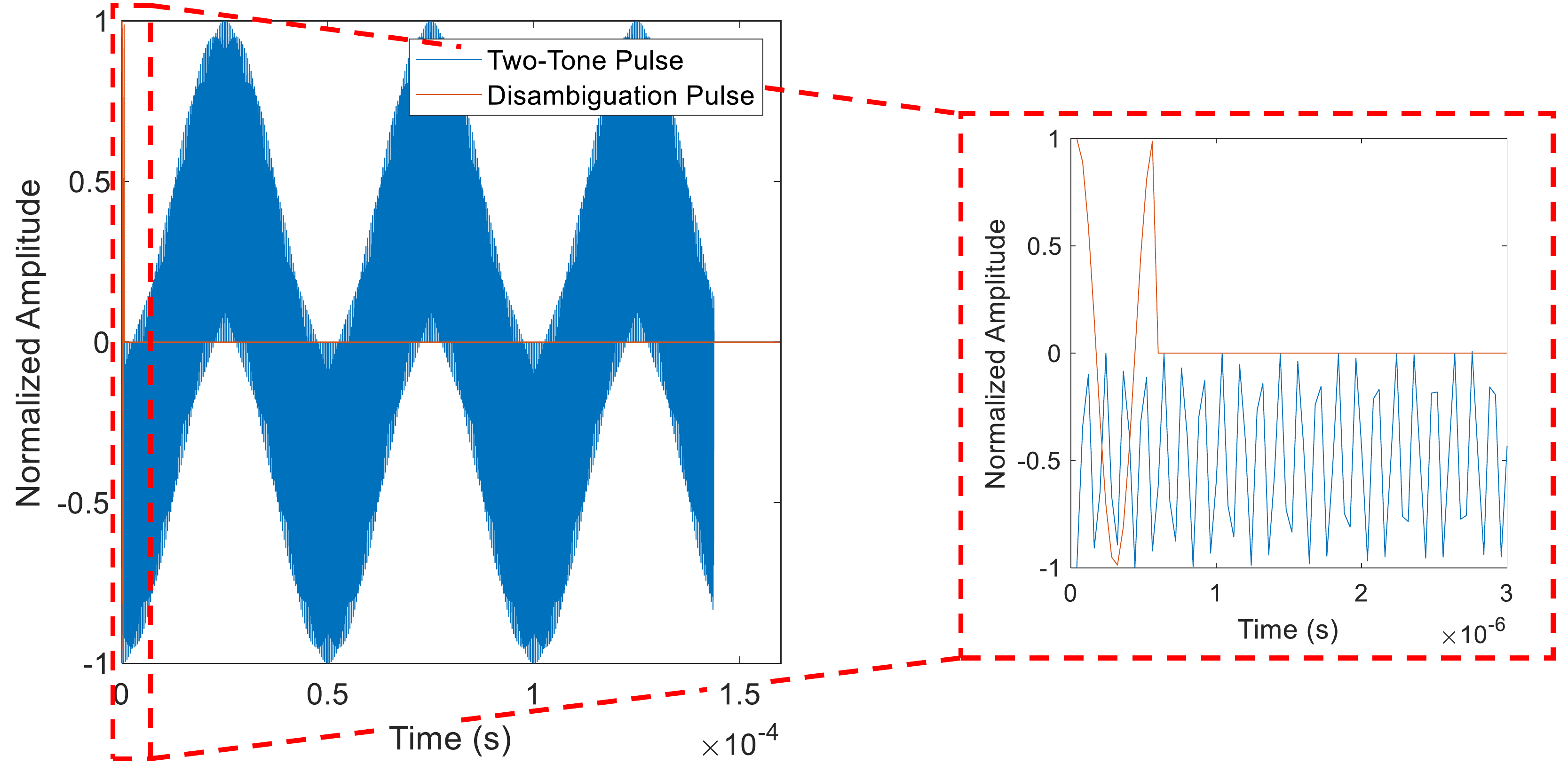}
	}%
	\\[0.2mm]%
	\subfloat[]{%
		\centering
		\includegraphics[width=\linewidth]{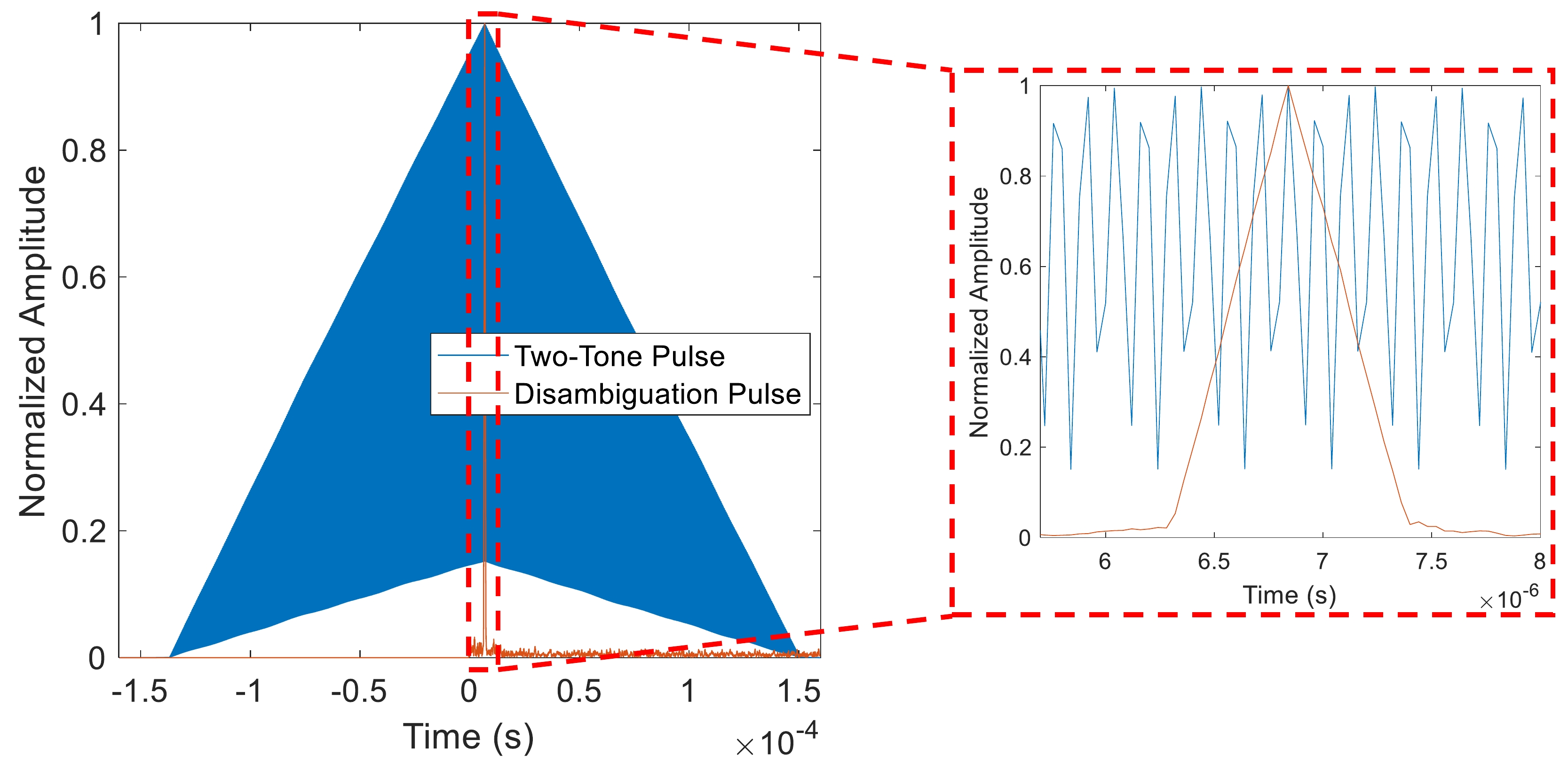}
	}%
	\\[2.6mm]
	\caption{(a) Two-tone ranging pulse with ${f_1=20}$~khz and ${f_2=7.52}$~MHz, along with a disambiguation pulse with ${f_d=1.875}$~MHz. These signals were sampled at 25 Msps. (b) Output of the matched filter for both the ranging pulse and the disambiguation pulse.}
	\label{fig:TTandDisamb}
\end{figure}

An example of the two-tone waveform along with a disambiguation pulse is shown in Fig.~\ref{fig:TTandDisamb}(a), where the two-tone frequencies were selected as ${f_1 = 20}$~kHz and ${f_2=7.52}$~MHz, while the disambiguation frequency was set to ${f_d=\delta f/2=1.875}$~MHz. The pulse width of the disambiguation pulse was 533~ns which is equal to one period, this will produce one lobe at the output of the matched filter; the disambiguation lobe will help in selecting the main lobe at the output of the matched filter of the two-tone signal as shown in Fig.~\ref{fig:TTandDisamb}(b).

\section{Wireless Frequency Synchronization}

Frequency synchronization is essential for coherent distributed arrays, whether they were operating in open-loop or closed-loop architectures. 
Conventionally, transceivers are frequency-synchronized via wired connections by connecting a reference frequency source to their internal phase-locked loops (PLLs). 
A self-mixing circuit 
designed for wireless frequency synchronization
was analyzed in \cite{8889331, SergeIMSBeamform}. The circuit 
captures a continuous two-tone signal transmitted from the primary node, producing a demodulated tone that is the difference frequency between the two tones. This tone separation is thus equal to the desired reference frequency, e.g. $f_m = 10$~MHz. The two-tone signal can either be the same as the ranging signal, or it can be a separate signal, providing certain frequency design rules are followed as discussed in \cite{9184602}.

The circuit operates by receiving a two-tone signal, splitting it into two identical paths, and mixing the two signals against one another in a single mixer. The normalized RF and LO inputs to the mixer can be given in general by
\begin{equation}\label{eq:VLO}
V_{RF} = \sin\left(2 \pi f_{s1} t + \phi_1\right) + \sin\left(2 \pi f_{s2} t + \phi_2\right),
\end{equation}
\begin{equation}\label{eq:VRF}
V_{LO} = \sin\left(2 \pi f_{s1} t + \phi_3\right) + \sin\left(2 \pi f_{s2} t + \phi_4\right),
\end{equation}
where $f_{s1}$ and $f_{s2}$ are the two synchronization tones, the phases $\phi_1$ and $\phi_2$ can be obtained once the separation between the two-tone transmitter and the circuit receiver has been determined, and the phases $\phi_3$ and $\phi_4$ are the phases $\phi_1$ and $\phi_2$ plus phase differences due to differing path lengths of the two signals to the RF and LO inputs. These paths are fixed, and can reasonably be designed to be path-matched, yielding $\phi_3 = \phi_1$ and $\phi_4 = \phi_2$. The resulting filtered mixer output is then
\begin{equation}\label{eq:VIF}
V_{IF} = \cos\left(2\pi\left(f_{s2}-f_{s1}\right)t+\phi_5\right),
\end{equation}
where $\phi_5 = \phi_2 - \phi_1$. The change in the phase constant $\phi_5$ can be tracked once the inter-node distance is estimated from the ranging system. 

\begin{figure}[t!]
	\centering
	\includegraphics[width=\linewidth]{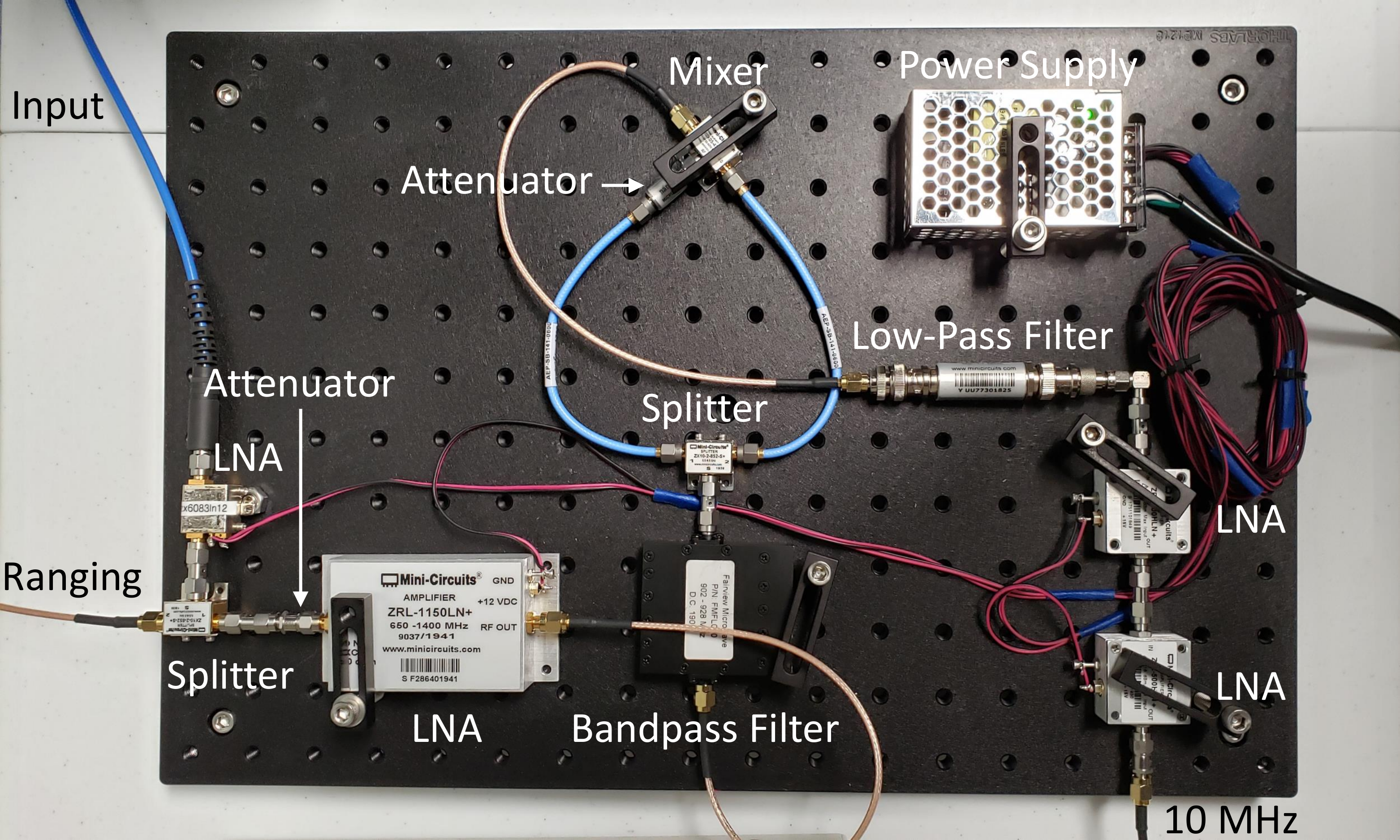}
	\caption[Optional caption]{Adjunct self-mixing frequency locking circuit that takes as input two-tone frequency synchronization signals to generate the reference frequency signals. The input ranging signals are filtered out and redirected to the appropriate input of the SDR.}
	\label{fig:FLC}
\end{figure}

The self-mixing frequency locking circuit used in this work is shown in Fig. \ref{fig:FLC}. 
The two-tone waveform is captured via antenna and amplified using a 20 dB RF low-noise amplifier (LNA) with 1.5 dB noise figure (NF). The signal is then split using a power divider, sending the signal to the transceiver to perform ranging on one path, and to the frequency synchronization circuit on the second path. The frequency synchronization signal is amplified using a 34 dB LNA with 0.8 NF, and the signals residing outside of the desired band are then filtered out using a cavity bandpass filter. A 7 dB attenuator is used before the LNA to prevent it from saturating in high-amplitude signal scenarios. The amplified signal is split and fed to the RF and LO inputs of a mixer with 6.2 dB conversion loss. The input to the RF port was attenuated by 10 dB to preserve the linear operation of the mixer. The mixer produces the desired 10 MHz frequency reference among other harmonics and unwanted frequencies. A low pass filter with 11 MHz cutoff frequency is used to keep the frequency of interest. Then, two 20 dB LNAs with 3.8 dB NF are used to drive the output power between 0 and 15 dBm, matching the input power requirements of the PLL on the SDR-based transceiver. This approach is directly scalable, as any node using a self-mixing circuit is capable of demodulating the frequency reference transmitted by the primary node.

\section{Long-Range Simultaneous Ranging and Frequency Synchronization}\label{sec:rangingAndFreq}

\begin{figure*}[t!]
	\centering
	\includegraphics[width=\linewidth]{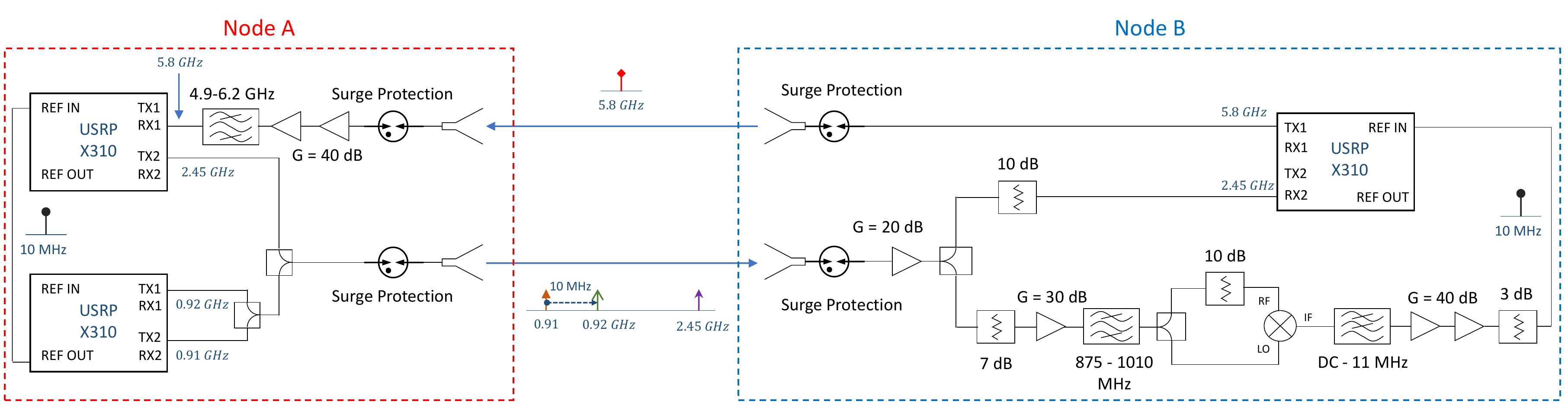}
	\caption[Optional caption]{Circuit diagram of the simultaneous frequency synchronization and range estimation. Node A is acting as a primary node while node B is the secondary one. Node A transmit the frequency synchronization signals at 910 MHz and 920 MHz and the ranging signal at a 2.45GHz carrier. Node B locks its frequency to node A and then amplifies and retransmits the ranging signals using a 5.8 GHz carrier \cite{SergeAPS2020LongRange}.}
	\label{fig:CircuitDiagram}
\end{figure*}

\begin{figure}[t!]
	\centering
	\includegraphics[width=0.9\linewidth]{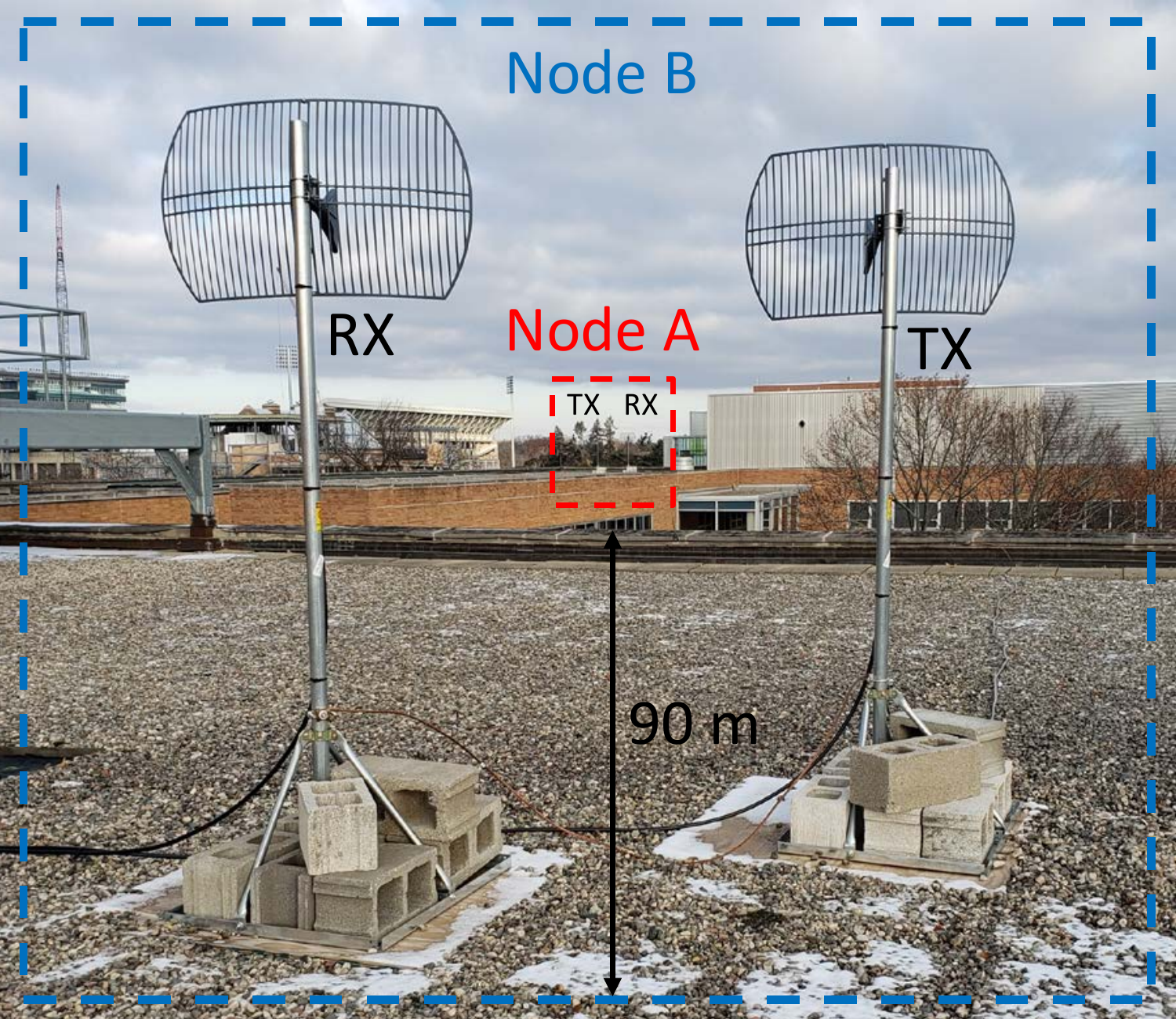}
	\caption[Optional caption]{Four Antennas were used in total for this experiment; one for transmit (TX) and one for receive (RX) at both nodes A and B. Nodes A and B were separated by 90 m.}
	\label{fig:Antennas}
\end{figure}

Wireless frequency synchronization is not only essential for enabling coherent beamforming of the transmitted signals, but also to allow cooperative ranging especially when the carrier of the transmitted signals from the repeater is different than the carrier of the received signals. In many cases, it is beneficial if the carrier frequencies for transmission and reception of the ranging signals are different to mitigate interference and multipath. Frequency synchronization is essential in this case since otherwise the downconversion and upconversion on the repeating node will introduce unknown phase errors.
This can be seen by considering the case where the radar transmits a baseband signal at a frequency $f_b$ modulated on a carrier $f_{c1}$. Once the repeating node receives this signal, it downconverts it using the carrier $f_{c1}^{'}$, which is nominally the same as the carrier $f_{c1}$ but in reality has drifted by some amount since the two nodes are not frequency locked. In this process, the frequency of the baseband signal at the secondary node becomes $f_b + f_{c1} - f_{c1}^{'}$. Afterwards, the secondary node retransmits back the signal at the second band using the carrier $f_{c2}^{'}$ which is the shifted version of the frequency $f_{c2}$. Similarly, once the signal is downconverted at the radar's end, the baseband frequency becomes $f_b + f_{c1} - f_{c1}^{'} + f_{c2}^{'} - f_{c2}$ which would ideally be equal to $f_b$. In order to recover the desired signal and perform range estimation, the two nodes need to be frequency locked in order to ensure that $f_{c1}^{'} = f_{c1}$ and $f_{c2}^{'} = f_{c2}$. Nevertheless, even if the same carrier was used for reception and retransmission ($f_{c2}^{'} = f_{c1}^{'}$), wireless frequency synchronization is still required for a significant reduction in the instantaneous frequency and phase errors.

\subsection{Experimental System}
Wireless frequency synchronization and range estimation were implemented in a SDR-based two-node system represented in the diagram in Fig. \ref{fig:CircuitDiagram}. Node A represents the primary node that was transmitting the frequency synchronization and ranging signals. Node B was the secondary node, which was locking its frequency to the frequency of the primary node using the circuit in Fig. \ref{fig:FLC}. Node B was also amplifying and retransmitting back the ranging signals that were being transmitted by node A. The frequency locking signals were continuous two-tone waveforms that were transmitted at 910~MHz and 920~MHz. The frequency synchronization frequencies were each generated on a separate daughterboard using an Ettus X310 SDR; the signals were only composed of the carrier frequencies in order to mitigate distortion from harmonics. The two-tone ranging signals were transmitted at a 2.45~GHz carrier. Node B first locked its oscillators to the reference frequency generated by the frequency locking circuit. Once locked, node B amplified the received signals and retransmitted them back using a 5.8 GHz carrier. Surge protection was installed between the antennas and the circuits, and the antennas were grounded. The two nodes were separated by 90~m as shown in Fig. \ref{fig:Antennas}. Wide band grid antennas operating from 600 MHz to 6.5 GHz were used, covering the entire operational bandwidth of the ranging and frequency transfer signals and had furthermore minimal wind resistance. The antennas gain was 15 to 26 dB, depending on the frequency band. This outdoor setup made it possible to observe the weather effects on the ranging performance. To monitor the weather conditions, a weather sensor was placed on the mast of the RX antenna of node A.

\subsection{Adaptive Control Architecture}

High ranging accuracy is essential for achieving high coherent gain. When ranging is performed in a well controlled environment, it might be possible to select a specific ranging bandwidth that will result in a desirable ranging standard deviation, which will allow a certain level of coherence for a desired carrier frequency. Selecting the desirable bandwidth can be accomplished through \eqref{eq:CRLB_r}, where for a specific post-processing SNR value, the minimum achievable ranging standard deviation can be calculated. However, performing range measurements in an outdoor environment introduces uncertainties regarding the SNR, interference level, antenna vibration, and others. These uncertainties makes it challenging to accurately estimate the ranging accuracy based on prior information and the bandwidth. To avoid complications, a much wider bandwidth than what is required can be used for ranging, ensuring that the desired ranging accuracy is always attained, however this results in unnecessary consumption of bandwidth potentially impacting spectral congestion and interference. Although two-tone waveforms are spectrally sparse, using a wider bandwidth could nonetheless introduce interference in unwanted bands. Thus, in an outdoor and dynamic node setup, it is important to use the narrowest bandwidth that will result in the appropriate accuracy.

\begin{figure}[t!]
	\centering
	\includegraphics[width=\linewidth]{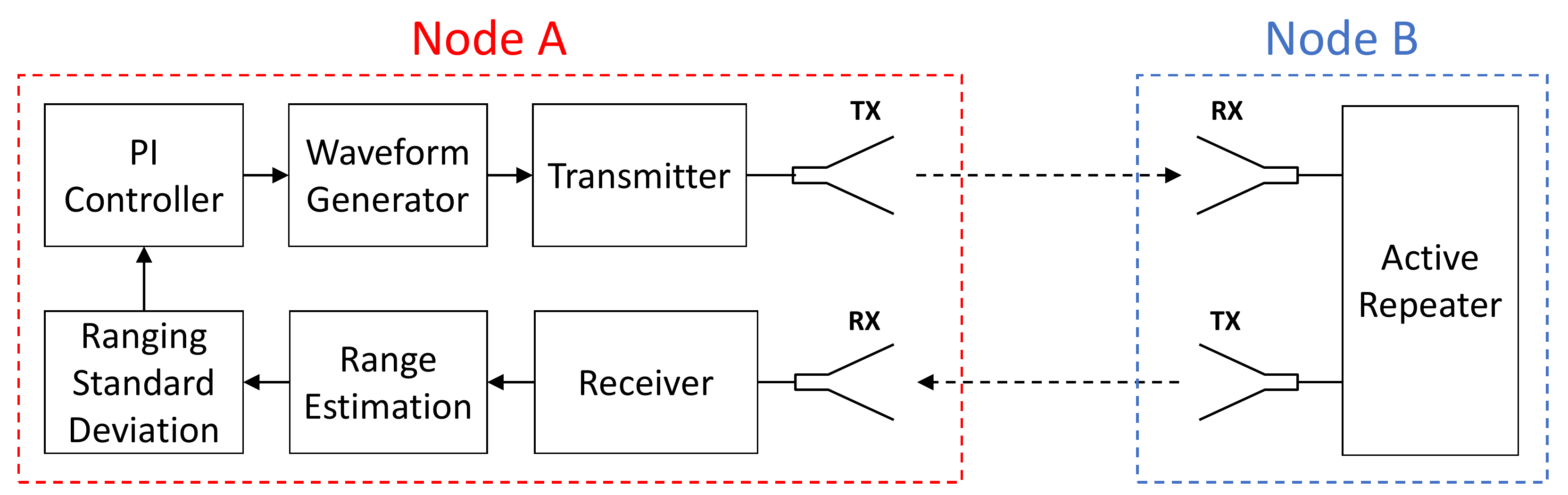}
	\caption[Optional caption]{Adaptive inter-node cooperative ranging architecture. Using this approach, node A modifies the bandwidth of its transmitted ranging waveform to counteract the environmental effects on the ranging accuracy, thus minimizing the bandwidth used to achieve coherent transmission from multiple nodes in the array.}
	\label{fig:AdaptiveLoop}
\end{figure}

To counteract the fluctuations in the ranging standard deviation and satisfy the dynamic bandwidth requirements, an adaptive processing architecture based on that from \cite{9115875} was developed to ensure that the system used only the minimum necessary bandwidth to obtain a desired range accuracy. Fig.~\ref{fig:AdaptiveLoop} summarizes the adaptive inter-node cooperative ranging architecture on the primary node A. During the measurement, node A transmits ranging pulses to node B, then node B amplifies the received pulses and retransmits them back to node A. 
Once multiple range estimates are obtained, the standard deviation of the range estimates is calculated; in our case, the ranging standard deviation was calculated using 200 consecutive range estimates. Afterwards, the difference between the estimated ranging standard deviation and the desired ranging standard deviation was considered as an error and this value was fed as an input to a PI controller. The PI controller was used to determine the appropriate tone separation to minimize the difference between the estimated and desired standard deviations. Finally, the new waveform was generated and transmitted to perform new range estimates. In this paper we refer to the duration of one full loop as a \textit{processing interval}.

Once the ranging standard deviation is obtained and the difference between the measured value and the reference point is calculated, a PI controller is used to minimize this difference by altering the transmitted waveform to counteract any increase or decrease in the ranging accuracy, ensuring that no more than the minimum sufficient bandwidth is consumed. The PI controller is derived from the proportional-integral-derivative (PID) controller, which is a widely used closed loop controller. 
PID controllers apply a desired correction based on proportional, integral, and derivative actions. The proportional term in the controller acts proportionally to the error in the ranging system, where the weight $K_p$ scales the magnitude of the action. The integral term, which is represented by the integration time $T_i$ is used to remove the offset at steady state. As the term indicates, the integral action takes into account both the magnitude of the error and the duration of that error, where the error is represented by the mismatch between the desired standard deviation and the actual one. Finally, the derivative term monitors the abrupt changes in the error and its main task is to minimize the overshoot in the system. The main drawback of using the derivative action is that in noisy and uncontrolled environments it can drive the system to instability, which led to disregarding this action in this work; hence the PI version of the PID controller was used. The discrete form a PI controller can be given by
\begin{equation}\label{eq:PI}
x[n] = x[n-1] + K_p \left[ \left(  1 + \frac{\Delta t}{T_i}  \right) e[n] -e[n-1]  \right],
\end{equation}
where the output of the controller $x[n]$ is inversely proportional to the tone separation of the two-tone ranging signal $2\delta f~=~x[n]$ (this is the case since as shown in \eqref{eq:CRLB_r}, the ranging standard deviation is inversely proportional to the tone separation), $e[n]$ is the error which represents the difference between the measured ranging standard deviation and the reference value, and $\Delta t = t_n - t_{n-1}$ is the time interval between two processing intervals.

The Ziegler-Nichols online tuning method \cite{ziegler1942optimum} a widely used approach for tuning PID or PI controllers, was used to determine the PI controller gains. The tuning is performed by first modifying only the proportional gain $K_p$ while setting the integration time $T_i$ to zero. In this process, $K_p$ is first set to zero and increased with small increments until reaching the ultimate gain $K_u$ which makes the system output oscillate with constant period and magnitude. The value of $K_u$ is recorded along with the oscillation period $T_u$. Finally the PI gains are set using 
\begin{equation}\label{eq:K_p}
K_p = 0.450~K_u,
\end{equation}
\begin{equation}\label{eq:T_i}
T_i = 0.833~T_u.
\end{equation}
Once these gains are assigned, the PI controller takes as an input the difference between the measured standard deviation and the desired value, referred to as the error $e[n]$. The output of the controller dictates the tone separation of the two transmitted ranging tones. The last step is to generate the waveform and use it for range estimation. Every time a specific number of new range estimates is obtained, which is in our case 200 estimates, the ranging standard deviation is recalculated and the appropriate controller action is taken.

\section{Experimental Results}


\subsection{Testing Design}
Experimental results were gathered using USRP X310 SDRs with two UBX 160 daughterboards per SDR.
All signals were transmitted on the Industrial, Scientific, and Medical (ISM) frequency bands as described in Section \ref{sec:rangingAndFreq}. The transmitted power was in accordance with the Federal Communcations Commission (FCC) regulations, where the maximum transmit power was less than 15 dBm, and due to the cable losses, the maximum Effective Isotropic Radiated Power (EIRP) was less than 36 dBm. The connection between the nodes and antennas was done using outdoor-rated cables LDF4-50A 1/2, where the cable length between every antenna and its SDR was equal to 20 m. The expected average attenuation caused by the cables for the frequency bands was as follows: 1.5 dB for the band 902 to 928 MHz, 2.4 dB for the band 2.4 to 2.5 GHz, and 4 dB for 5.725 to 5.875 GHz. For the antenna gain at these bands, the estimated values were as follows: 16 dBi for the band 902 to 928 MHz, 22 dBi for the band 2.4 to 2.5 GHz, and 25 dBi for 5.725 to 5.875 GHz. As for the antenna placement, the antennas were placed in pairs separated by 2 m, and the separation between the antennas of nodes A and B was equal to 90 m as shown in Fig. \ref{fig:Antennas}.

Ranging signals were transmitted at a 2.45 GHz carrier with sideband signals modulated with frequency offsets ${f_1 = 20}$~kHz and 20~kHz~${\leq f_2 \leq}$~7.52~MHz. The upper frequency $f_2$ was controlled adaptively using the PI controller, where the operational range of the tone separation was between 0 and 7.5~MHz. The disambiguation pulse consisted of one tone at 1.875~MHz. The ranging pulse width was equal to $143.7~\mu$s while the disambiguation pulse width was equal to 533~ns. The pulse repetition interval was set to $159.7 ~\mu$s for both the ranging and disambiguation pulses. Every ranging cycle included the transmission of one ranging pulse along with one disambiguation pulse and the reception of these pulses along with the required signal processing procedure. The duration for every ranging cycle was equal to $319.4~\mu$s. When conducting the experiments, ranging and disambiguation pulses were transmitted every  105~ms to allow data logging while preventing any buffer underflow or overflow. All the signal processing was performed in real-time using LabVIEW.


The targeted ranging standard deviation for the adaptive controller was 10 mm.
The ranging standard deviation was calculated using 200 sequential pulses; the pulses were assembled in 40 groups of 5 consecutive elements each, and then these 5 consecutive samples in every group were averaged before estimating the standard deviation. This procedure reduces the calculated ranging standard deviation by a factor of $\sqrt{5}$. Every time the standard deviation is estimated, the PI controller takes an action. The values of the controller gains were $K_p = 10 \mu$ and $T_i = 3.3$.

\subsection{Weather Effects on the Outdoor Ranging Accuracy} \label{sec:weatherEffect}

\begin{figure}[t!]
	\centering
	\centering%
	\subfloat[]{%
		\centering
		\includegraphics[width=0.49\textwidth]{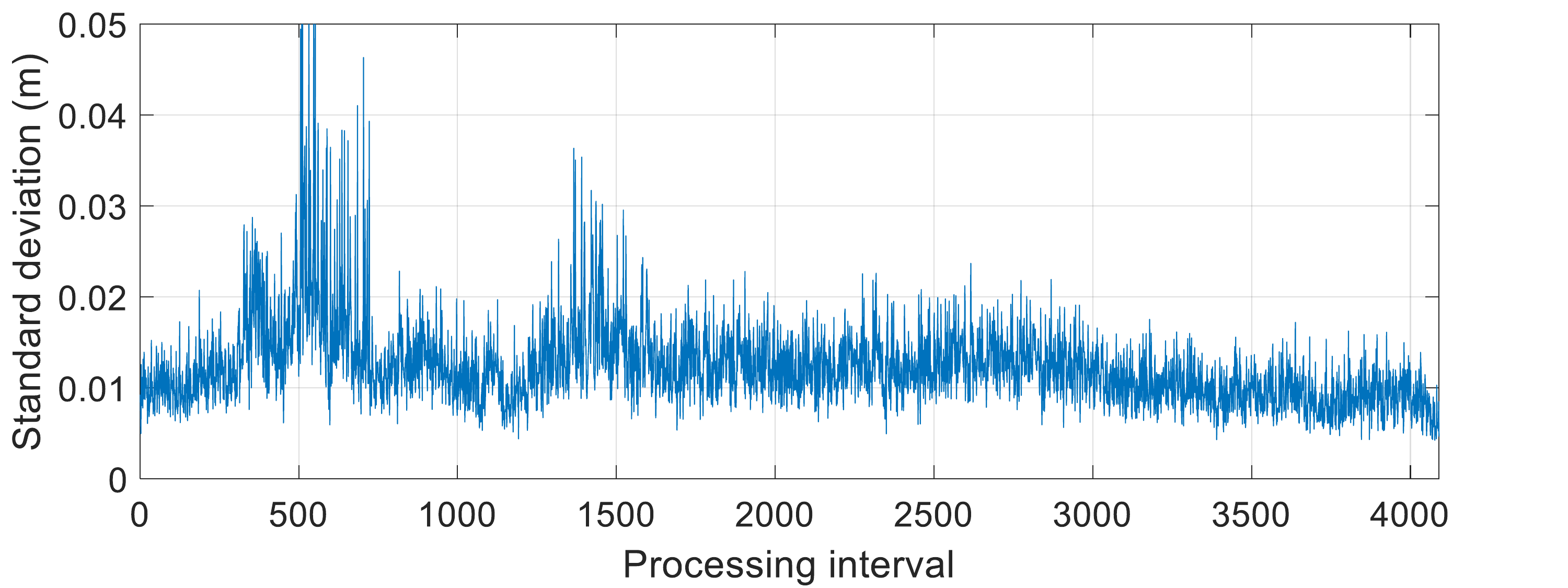}
	}%
	\\[0.2mm]%
	\subfloat[]{%
		\centering
		\includegraphics[width=0.49\textwidth]{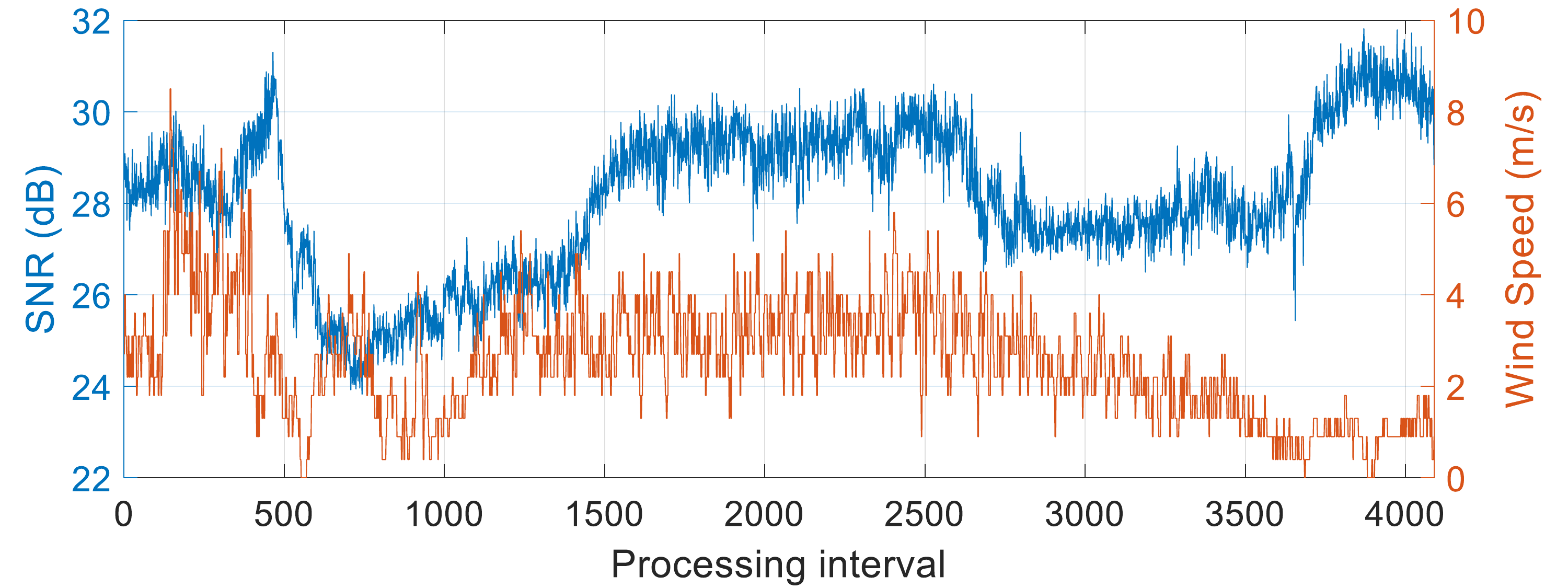}
	}%
	\\[0.2mm]%
	\subfloat[]{%
		\centering
		\includegraphics[width=0.49\textwidth]{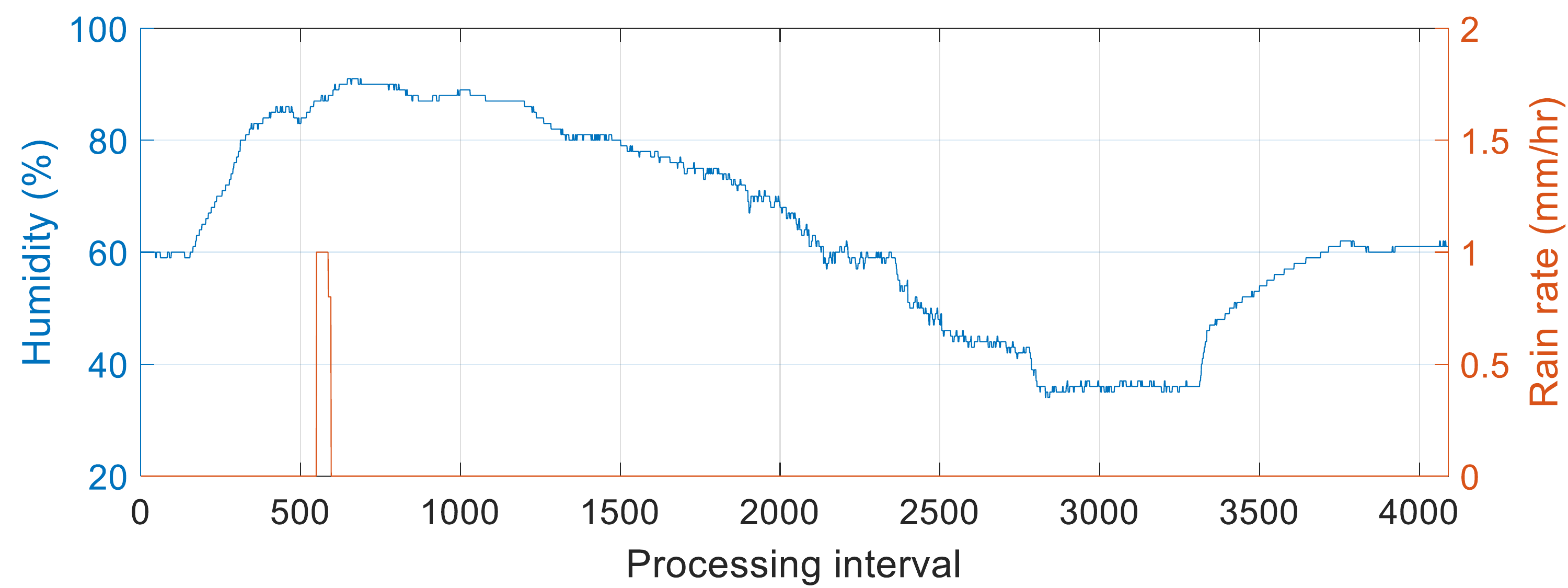}
	}%
	\\[2.6mm]
	\caption{(a) Ranging standard deviation for a 24 hours interval, where ${f_1 = 20}$~kHz and ${f_2 = 3.5}$~MHz. The effects of SNR changes and wind speed were shown in (b). Also the contribution of humidity and rain rate was presented in (c).}
	\label{fig:rangingSTDnoPI}
\end{figure}

Weather conditions were recorded using the weather sensor Davis 6250 Vantage Vue, which was attached to the mast of the RX antenna at node A. Using this weather sensor, it was possible to monitor the humidity, rainfall, wind speed, wind direction, pressure, and temperature. Weather data was collected with a logging interval of 1 minute, and it was used to analyze the effect of various weather conditions on the ranging accuracy.
The effects of weather conditions on the ranging standard deviation were recorded for 24 hours in total from June 1, 2020 23:56 until June 2, 2020 23:56. 4090 processing intervals were present in the 24 hours interval. In this test the adaptive framework was not used and $f_2$ was set to 3.5 MHz. The ranging standard deviation is shown in Fig. \ref{fig:rangingSTDnoPI}(a). It can be seen that although the tone separation was kept constant, the ranging standard deviation was varying due to various weather conditions and due to changing interference levels. In general, the high wind speeds lead to high antenna vibrations which negatively impacted the ranging accuracy. The humidity levels and rain rates increased the attenuation of the transferred signals, causing the SNR to drop along with the ranging accuracy. Furthermore, interference mainly caused by surrounding WiFi networks that were located close to the setup and were transmitting either on the same or neighboring channels impacted the SNR.

As shown, in Figs. \ref{fig:rangingSTDnoPI}(b) and \ref{fig:rangingSTDnoPI}(c), the drop in SNR from 550 to 1600 had a negative impact on the ranging standard deviation, and high wind speeds around 300 processing interval had a negative impact; once the wind speed decreased below 3 m/s starting from 3000 processing intervals, the ranging standard deviation decreased and started fluctuating around 0.01 m. Beyond the processing interval 3700, the average standard deviation of the ranging accuracy dropped below 10 mm due to the high SNR levels and low wind speeds. Also, high humidity percentages and rain drops contributed to the increase in the standard deviation of ranging around the processing interval 700.

Using the results from Fig. \ref{fig:rangingReq}, the maximum achievable coherent beamforming frequencies were calculated for the 2 nodes case. Three probabilities for achieving coherent gain of at least 0.9 were investigated: 0.9, 0.8, and 0.7. To improve the clarity of the figure, the 4090 individual estimates for each probability were plotted with fading colors, while their smoothed outputs, which were generated using a moving 10-point average window, were plotted with solid colors. As shown, without an adaptive ranging approach, the maximum achievable beamforming frequency varies from below 1 GHz to 2 GHz for the $P=0.9$ case, with commensurately higher frequencies for lower probabilities. It is thus not possible to select one value for the tone separation that could achieve the appropriate ranging accuracy for the desired coherent frequency while minimizing the waveform bandwidth.

\begin{figure}[t!]
	\centering
	\includegraphics[width=\linewidth]{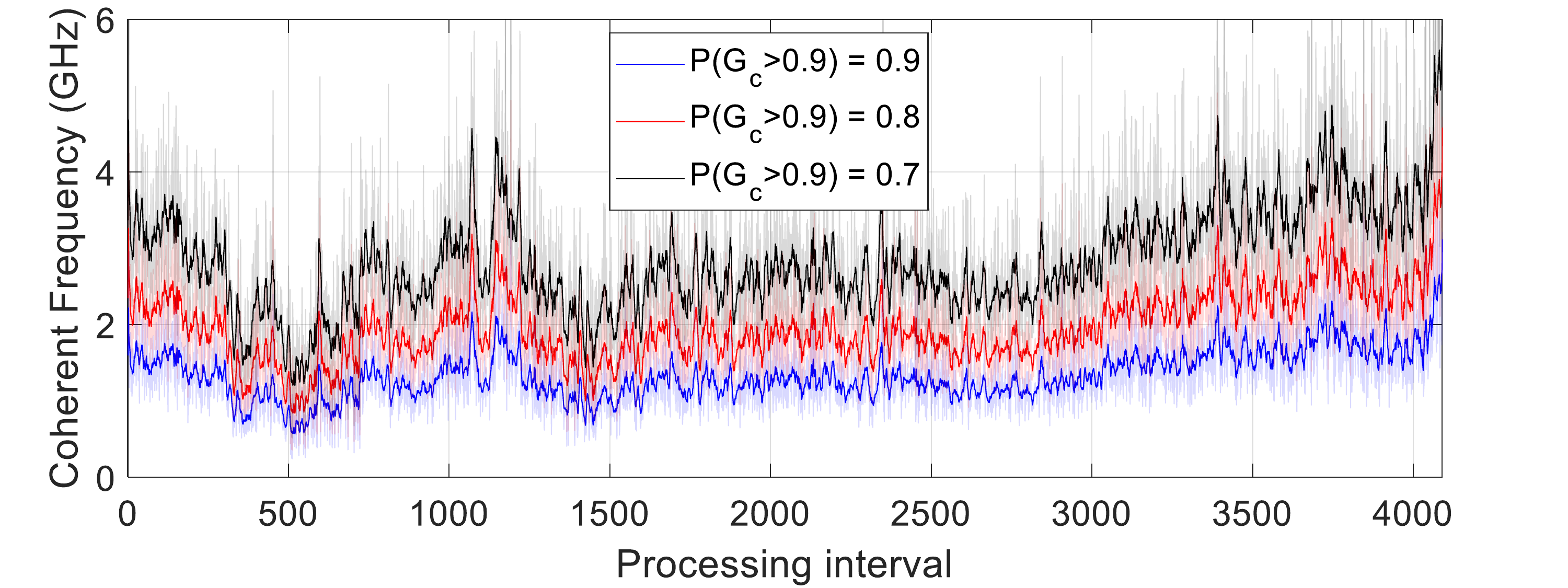}
	\caption[Optional caption]{Achievable coherent frequency using the obtained ranging standard deviation. Three probabilities for achieving coherent gain of at least 0.9 were investigated. The lines with solid colors represent the moving average with a window of 10, while the faded lines represent the individual estimates.}
	\label{fig:GcnoPI}
\end{figure}

\subsection{Adaptive Ranging System} \label{sec:PIWeather}

\begin{figure}[t!]
	\centering
	\centering%
	\subfloat[]{%
		\centering
		\includegraphics[width=0.49\textwidth]{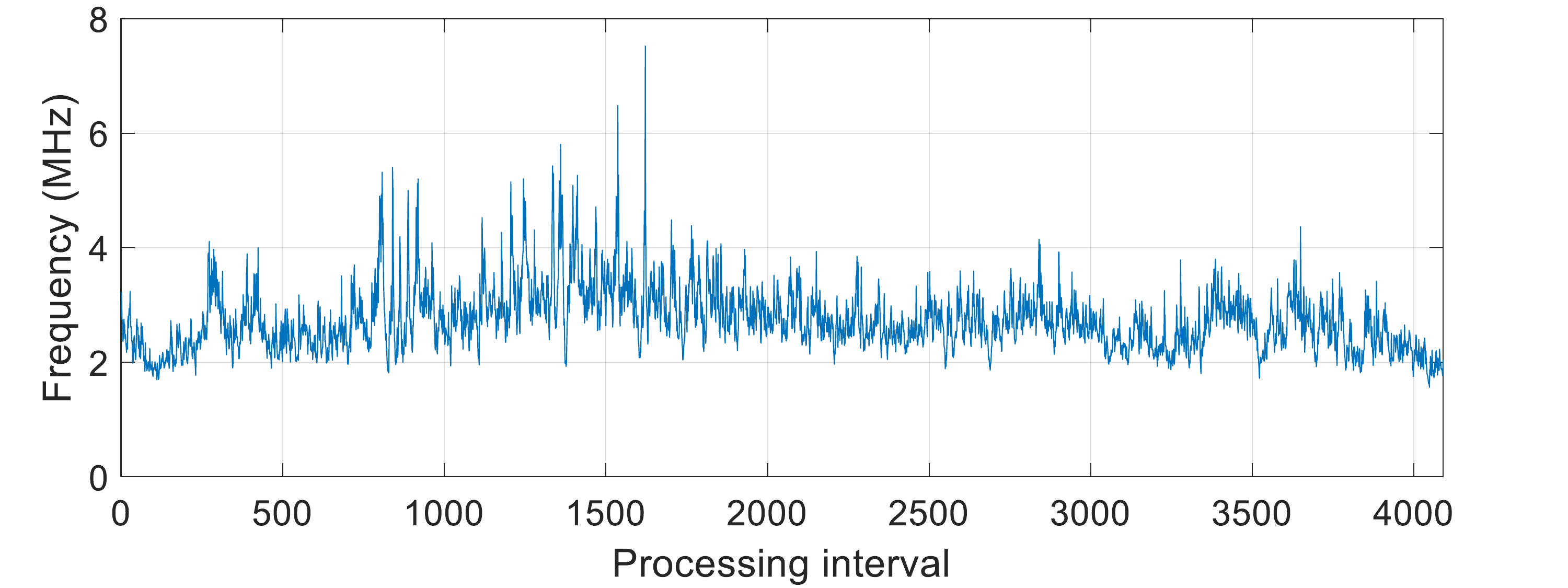}
	}%
	\\[0.2mm]%
	\subfloat[]{%
		\centering
		\includegraphics[width=0.49\textwidth]{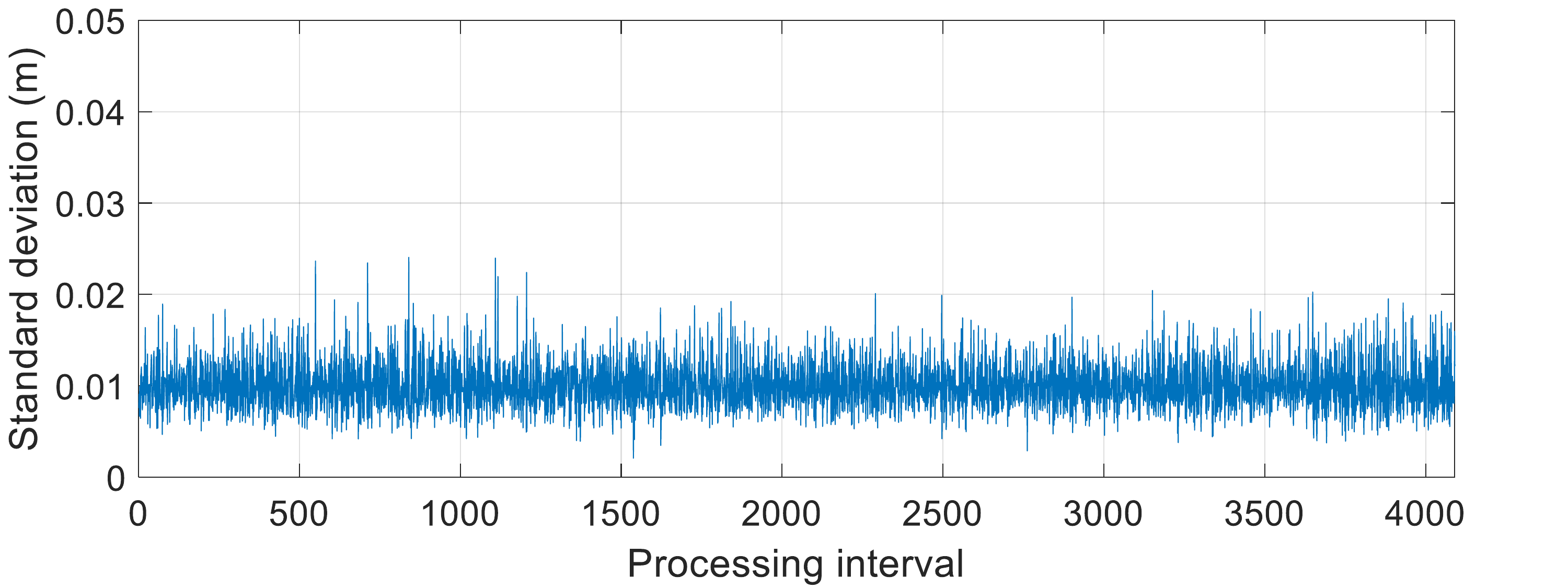}
	}%
	\\[0.2mm]%
	\subfloat[]{%
		\centering
		\includegraphics[width=0.49\textwidth]{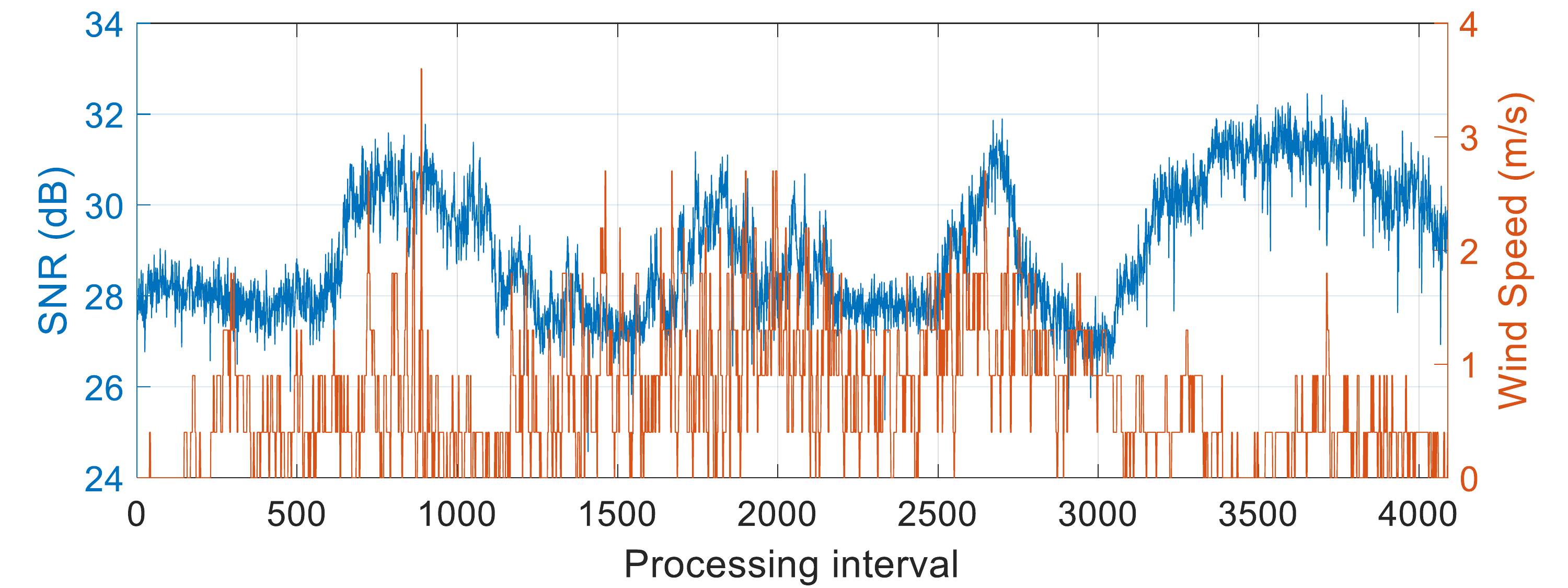}
	}%
	\\[0.2mm]%
	\subfloat[]{%
		\centering
		\includegraphics[width=0.49\textwidth]{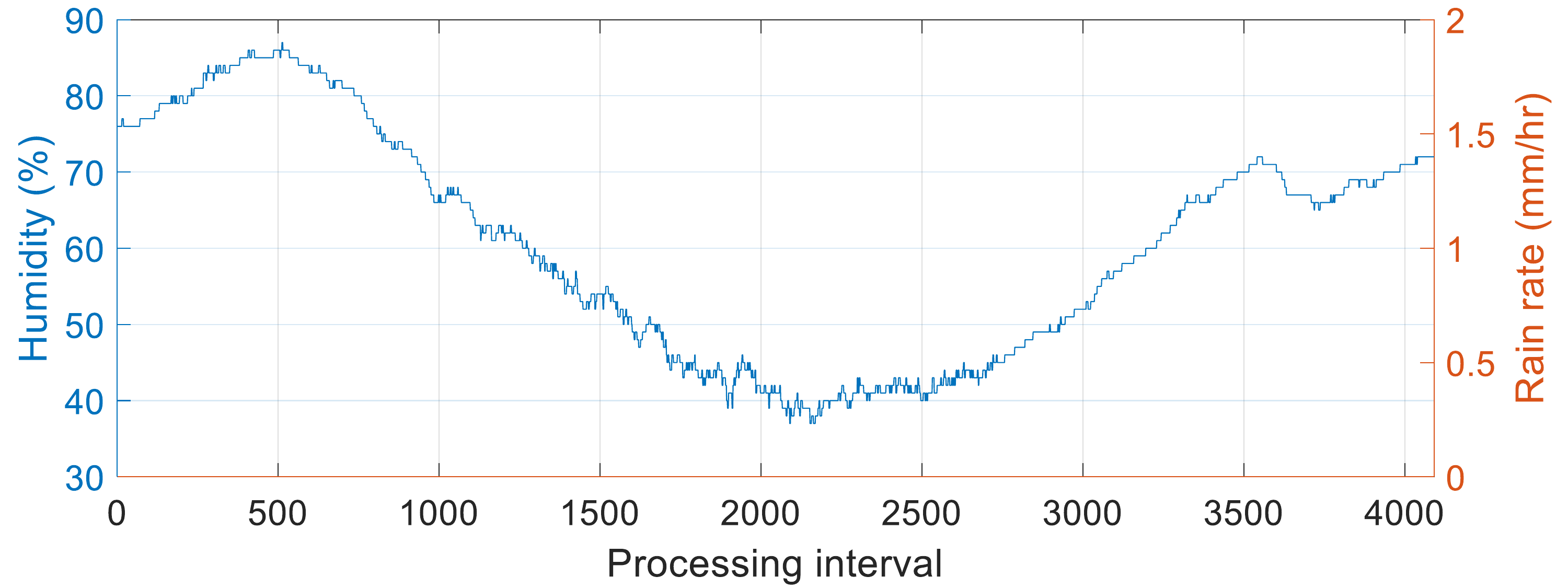}
	}%
	\\[2.6mm]
	\caption{(a) The adaptively modified second tone $f_2$ of the transmitted ranging pulses. (b) Ranging standard deviation for a 24 hours interval, where ${f_1 = 20}$~kHz and $f_2$ was adaptively modified. The effects of SNR changes and wind speed were shown in (b). Also the contribution of humidity and rain rate was presented in (c).}
	\label{fig:rangingSTDwithPI}
\end{figure}

\begin{figure}[t!]
	\centering
	\includegraphics[width=\linewidth]{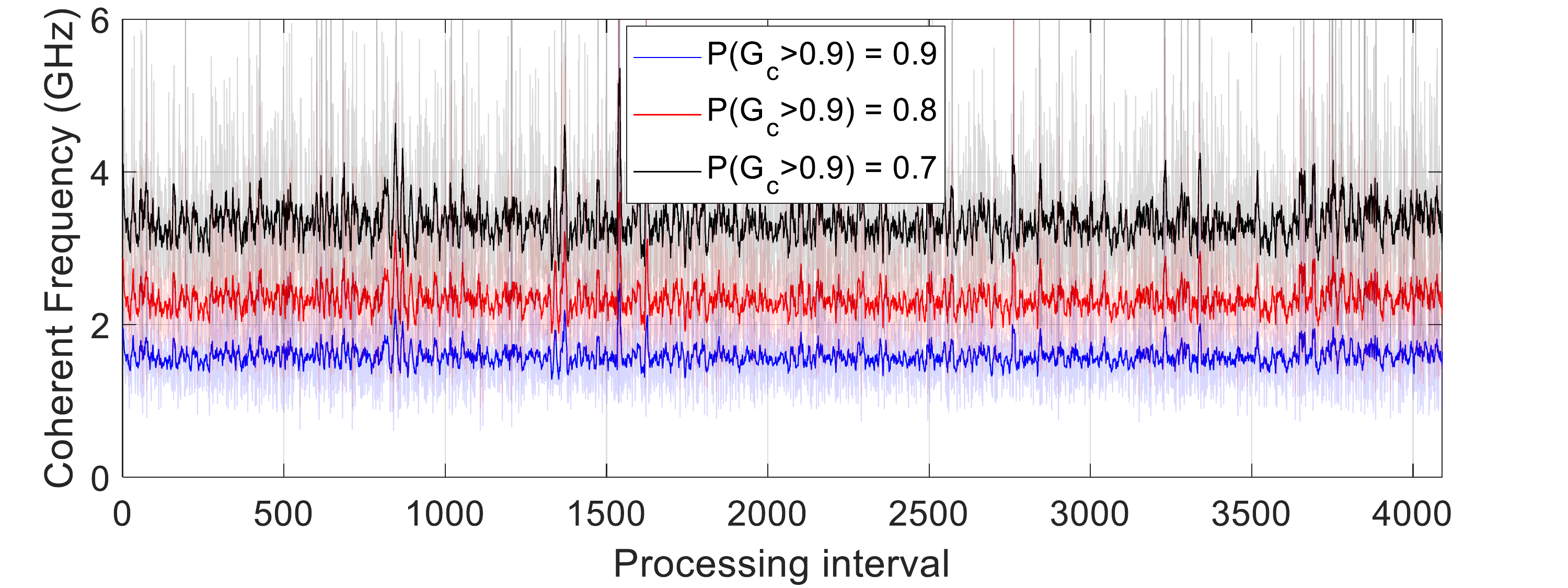}
	\caption[Optional caption]{Achievable coherent frequency using the obtained ranging standard deviation in the adaptive framework. Three probabilities for achieving coherent gain of at least 0.9 were investigated. The lines with solid colors represent the moving average with a window of 10, while the faded lines represent the individual estimates.}
	\label{fig:GcwithPI}
\end{figure}

The experiment was conducted a second time using the adaptive ranging framework, and the second tone was adaptively modified to maintain a ranging accuracy close to 10~mm. The effects of weather conditions on the ranging standard deviation were recorded for 24 hours in total from June 3, 2020 03:05 until June 4, 2020 03:05. Fig. \ref{fig:rangingSTDwithPI} shows the second tone frequency $f_2$, along with the ranging standard deviation, SNR, wind speed, humidity, and rain rate. As shown, $f_2$ was adaptively modified to counteract the various effects caused by the varying weather conditions. It was possible to use the minimum bandwidth for this ranging waveform in order to achieve the desired ranging accuracy.

In this case, since $f_2$ was modifying adaptively to maintain the desired ranging standard deviation, we can see the effect of the SNR and weather conditions on the transmitted ranging frequencies. The variations in SNR and wind speed had the biggest contributions; whenever the SNR dropped and the wind speed increased, we were able to see an increase in $f_2$.
Using Fig. \ref{fig:rangingReq}, the achievable coherent beamforming frequencies were recalculated for the case where we used the adaptive framework, as we can see in Fig. \ref{fig:GcwithPI}, it was possible to maintain a specific coherent beamforming frequency using this approach. For instance, by setting the ranging standard deviation to 10 mm, it is possible to maintain a coherent action at frequencies 1.5, 2.2, and 3 GHz for the probabilities 0.9, 0.8, and 0.7 to achieve at least 0.9 coherent gain.

\begin{figure*}[t!]
	\centering
	\centering%
	\subfloat[]{%
		\centering
		\includegraphics[width=\linewidth]{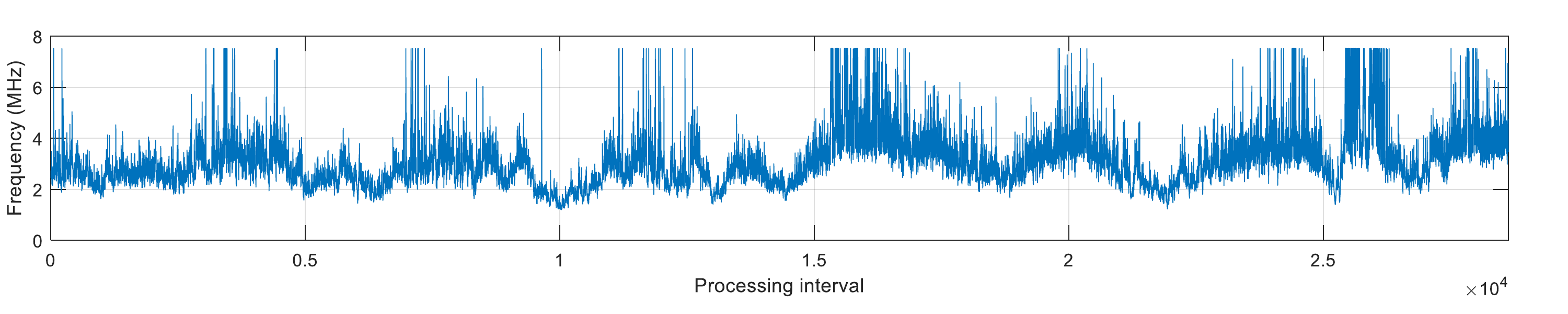}
	}%
	\\[0.2mm]%
	\subfloat[]{%
		\centering
		\includegraphics[width=\linewidth]{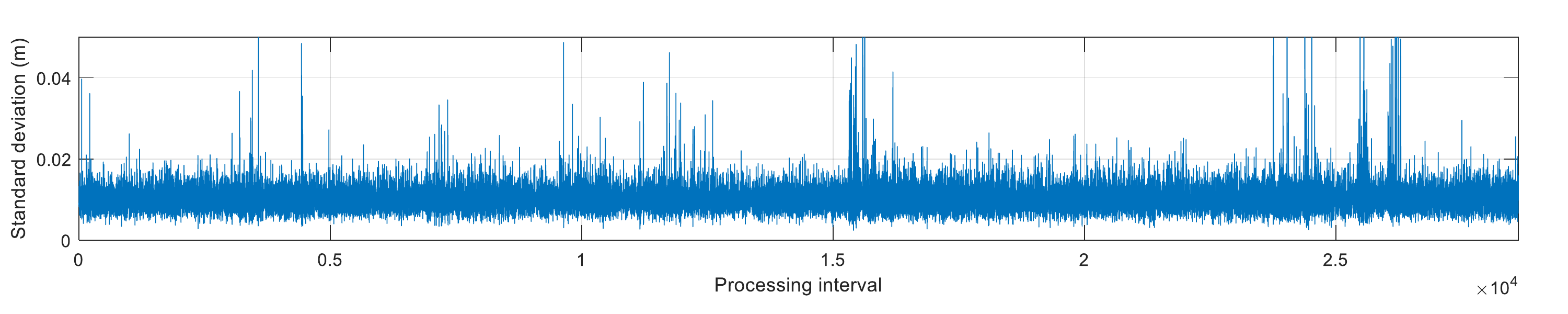}
	}%
	\\[0.2mm]%
	\subfloat[]{%
		\centering
		\includegraphics[width=\linewidth]{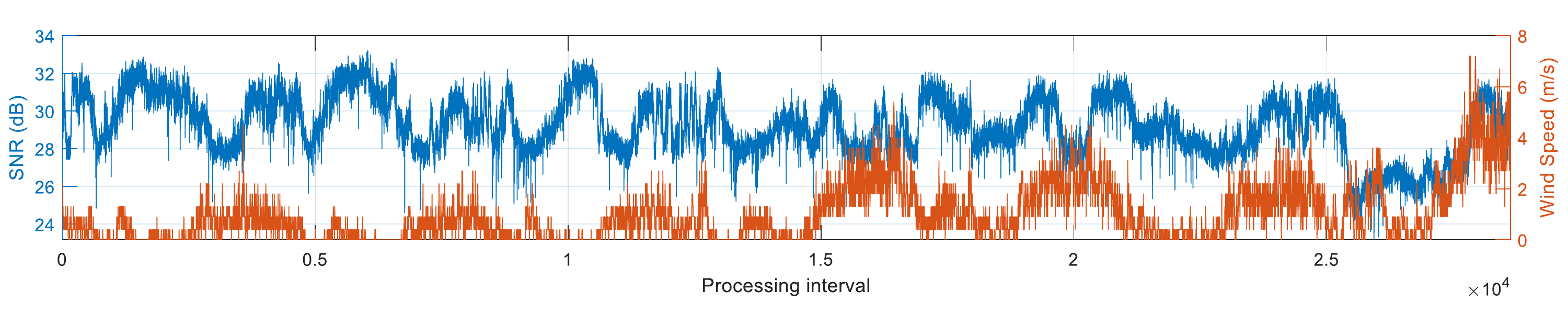}
	}%
	\\[0.2mm]%
	\subfloat[]{%
		\centering
		\includegraphics[width=\linewidth]{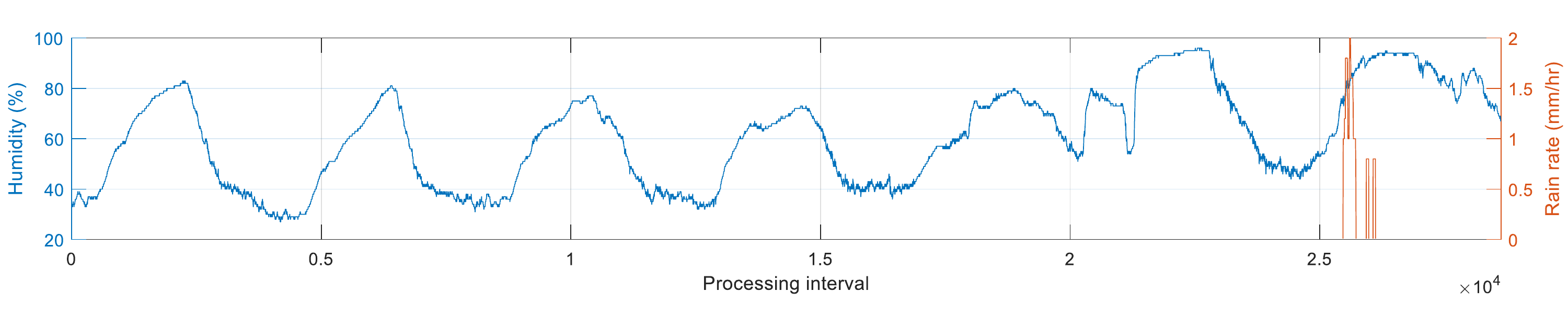}
	}%
	\\[2.6mm]
	\caption{(a) The adaptively modified second tone $f_2$ of the transmitted ranging pulses. (b) Ranging standard deviation for a 1 week interval, where ${f_1 = 20}$~kHz and $f_2$ was adaptively modified. The effects of SNR changes and wind speed were shown in (b). Also the contribution of humidity and rain rate was presented in (c).}
	\label{fig:rangingSTDwithPI7}
\end{figure*}

\begin{figure*}[t!]
	\centering
	\includegraphics[width=\linewidth]{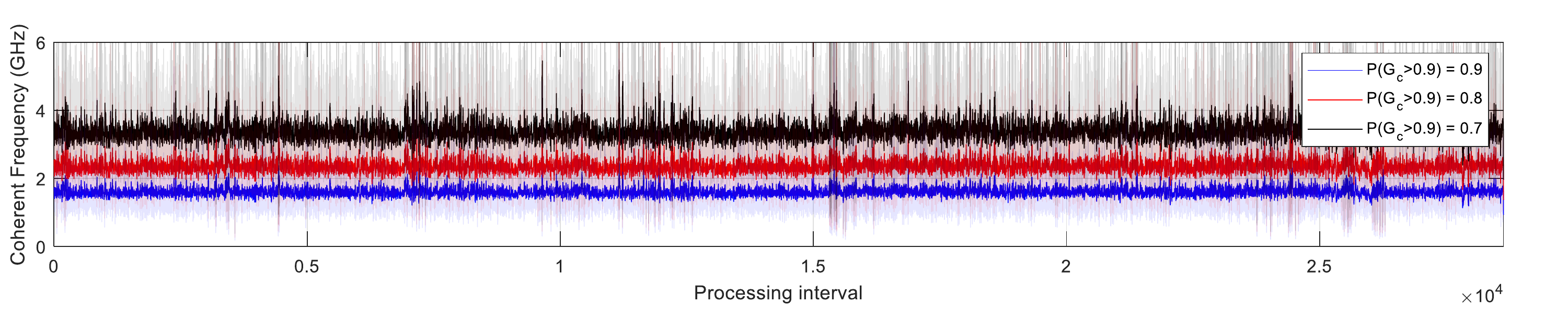}
	\caption[Optional caption]{Achievable coherent frequency using the obtained ranging standard deviation in the adaptive framework. Three probabilities for achieving coherent gain of at least 0.9 were investigated. The lines with solid colors represent the moving average with a window of 10, while the faded lines represent the individual estimates.}
	\label{fig:GcwithPI7}
\end{figure*}

The same experiment was then performed for a longer period. The effects of SNR fluctuations and weather conditions on the ranging standard deviation were recorded for 7 days in total from June 16, 2020 16:53 until June 23, 2020 16:53. Fig. \ref{fig:rangingSTDwithPI7} shows the second tone frequency $f_2$, along with the ranging standard deviation, SNR, wind speed, humidity, and rain rate. As shown, $f_2$ was adaptively modified to counteract the various effects caused by the varying weather conditions for the entire week. 28,630 processing intervals were captured in total. The achievable coherent beamforming frequencies were calculated for the case where we used the adaptive framework for 1 full week, the results are shown in Fig. \ref{fig:GcwithPI7}. By setting the ranging standard deviation to 10 mm, it is possible to maintain a coherent action at frequencies 1.5, 2.2, and 3~GHz for the probabilities 0.9, 0.8, and 0.7 to achieve at least 0.9 coherent gain.

\section{Conclusion}
An adaptive, long-range wireless phase synchronization approach for distributed microwave transceivers was demonstrated in this work. Cooperative ranging using two different channels for transmit and receive was feasible once the two separated nodes were frequency locked using an adjunct self-mixing circuit to demodulate a frequency reference. Synchronization was demonstrated outdoors over a 90~m separation; this distance can be extended up to kilometers as long as the SNR requirements on the self-mixing receiver are ensured. The adaptive ranging framework was based on a PI controller, where the tone separation of the two-tone ranging signal was modified adaptively to ensure that a reference ranging standard deviation, which was in our case 10 mm, is achieved. Due to unpredictable interference levels and weather conditions, such approach is necessary to ensure high coherent gain for certain carrier frequencies while minimizing the bandwidth or tone separation that was used by the ranging pulses. In our work we interpreted the effects of ranging standard deviation on the coherent gain for open-loop coherent distributed arrays; we evaluated the ability to achieve 90\% coherent gain. Using 10 mm as a reference standard deviation for ranging, it was shown that 0.9 coherent gain for carrier frequencies up to 1.5, 2.2, and 3 GHz can be achieved for the probabilities 0.9, 0.8, and 0.7 respectively.




\bibliographystyle{IEEEtran}

\end{document}